\documentclass[aps,prb,showpacs,twocolumn,superscriptaddress,amsmath,amssymb,floatfix,longbibliography]{revtex4-1}

\usepackage{graphicx}
\usepackage{float}
\usepackage{latexsym,amsmath,amssymb,bm,euscript}
\usepackage{color}
\usepackage{wasysym}
\usepackage{epstopdf}
\usepackage[colorlinks=True,linkcolor=blue,urlcolor=blue,citecolor=blue]{hyperref}
\usepackage{type1cm}
\usepackage{appendix}
\usepackage{mathtools}

\usepackage{amsfonts, amssymb}
\usepackage{stackengine}
\usepackage{scalerel}
\usepackage{xcolor}
\usepackage{graphicx}
\usepackage{CJK}

\hypersetup{
           breaklinks=true,
           colorlinks=true,
           pdfusetitle=true,
        }

\begin{document}

\begin{CJK*}{GBK}{}

\title{The principle of learning sign rules by neural networks in qubit lattice models}
\author{Jin Cao}
\affiliation {Department of Physics and Center for Advanced Quantum Studies, Beijing Normal University, Beijing 100875, China}
\author{Shijie Hu}
\email{shijiehu@csrc.ac.cn}
\affiliation{Beijing Computational Science Research Center, Beijing 100193, China}
\affiliation{Department of Physics, Beijing Normal University, Beijing 100875, China}
\author {Zhiping Yin}
\affiliation {Department of Physics and Center for Advanced Quantum Studies, Beijing Normal University, Beijing 100875, China}
\author{Ke Xia}
\affiliation{Department of Physics, Southeast University, Nanjing 211189, China}

\date{\today}

\begin{abstract}
A neural network is a powerful tool that can uncover hidden laws beyond human intuition. However, it often appears as a black box due to its complicated nonlinear structures. By drawing upon the Gutzwiller mean-field theory, we can showcase a principle of sign rules for ordered states in qubit lattice models. We introduce a shallow feed-forward neural network with a single hidden neuron to present these sign rules. We conduct systematical benchmarks in various models, including the generalized Ising, spin-$1/2$ XY, (frustrated) Heisenberg rings, triangular XY antiferromagnet on a torus, and the Fermi-Hubbard ring at an arbitrary filling. These benchmarks show that all the leading-order sign rule characteristics can be visualized in classical forms, such as pitch angles. Besides, quantum fluctuations can result in an imperfect accuracy rate quantitatively.
\end{abstract}

\maketitle
\end{CJK*}

Decoding hidden information from the ground-state wave function is essential for understanding the properties of quantum closed systems at zero temperature, including orders, correlations, and even intricate entanglement features, etc~\cite{Eisert2010, Chertkov2018, Wang2019, Irkhin2019}. For a real Hamiltonian, the sign structure of elements in the real wave function can be summarized as a {\it sign rule} within a selected representation~\cite{Grover2015}. For example, the Perron-Frobenius theorem is applied to a class of Hamiltonians with non-positive off-diagonal elements~\cite{Perron1907, Frobenius1909}. The Marshall-Peierls rule (MPR) is another example applicable to antiferromagnetic spin models on bipartite lattices~\cite{Marshall1955,  Lieb1961, Lieb1962}. These sign rules have been believed to be connected to various physical phenomena, such as the volume law for the R\'{e}nyi entanglement entropies~\cite{Grover2015}, spatial periodicity of states~\cite{Zeng1995, Bursill1995}, phase transitions~\cite{Retzla1993, Richte1994, Cai2018, Westerhout2020}, and so on.

Similar to the matrix product state (MPS) successfully applied to (quasi) one-dimensional (1D) lattice models~\cite{White1992, Wang1999, Schollwock2011, Orus2014}, the neural network quantum state (NNQS) and fast-developing machine learning (ML) techniques provide a new approach for multi-scale compression of the wave function, which has been widely used in one and higher dimensional quantum many-body systems~\cite{Carleo2017, Carleo2019, Jia2019, Vivas2022}. By using the empirical activation function {\it cosine} in the hidden layer of NNQS, the complicated sign rules in qubit lattice models can be learned from the wave functions~\cite{Cai2018}, and subsequent studies have drawn significant attention in recent years~\cite{Choo2019, Westerhout2020, Szabo2020, Bukov2021}. These studies have shown that it is a practical advantage to enhance the representation precision for complex sign rules~\cite{Cai2018, Choo2019, Westerhout2020, Szabo2020, Bukov2021}. This can be achieved by adding more hidden layers/neurons in NNQS or designing new architectures. Meanwhile, there is a growing concern about interpreting the meaning of highly nonlinear structures in neural networks~\cite{Roscher2020, He2020, Fan2021} and finding links to existing physical insights~\cite{Raissi2019, Yuan2021, Cai2021}, which strongly motivates our work.

In this work, we establish a Gutzwiller mean-field (GWMF) principle of the sign rules for ordered ground states in qubit lattice models. The leading-order term can be well understood using a single-hidden-neuron feed-forward neural network (shn-FNN). Our findings, tested on various spin and fermion models, suggest that the leading-order sign rules have clear physical interpretations tightly related to orders in spins or charges. The structure of the paper is organized as follows: In Sec.~\ref{secA}, we present the GWMF picture of the sign rules for ordered states in qubit lattice models. In Sec.~\ref{secB}, we introduce shn-FNN in detail, which matches the GWMF picture and can be easily interpreted. In Sec.~\ref{secC}, we demonstrate the technical details of data set preparation and shn-FNN training. In Sec.~\ref{secD}, we apply shn-FNN to extract the leading-order sign rules in spin and Fermi-Hubbard models. We also discuss the influence of frustration and global symmetries. At last, we summarize conclusions and make a discussion briefly in Sec.~\ref{secE}.

\section{Gutzwiller mean-field theory} \label{secA}
A qubit is commonly used to represent a quantum state $\rvert n \rangle$ in various fields of condensed matter physics. Examples include a spin-$1/2$ in quantum magnets~\cite{Vasiliev2018}, a single fermion state in ultra-cold atomic systems~\cite{Gardiner2017}, a two-level atom in quantum cavities~\cite{Meher2022}, and so on~\cite{Makhlin2001, Luo2008, Kjaergaard2020}. The binary value $n$ corresponds to an empty or occupied fermion level, or a spin-$1/2$ polarizing $\uparrow$ or $\downarrow$ in the z-axis. For a lattice with $L$ qubits, the basis can be expressed as $\rvert \mathbf{n} \rangle = \otimes^{L}_{l=1} \rvert n_l \rangle$, where $\rvert n_l \rangle$ represents the local basis at site-$l$, and the quantum indices $n_l=0$, $1$ form a vector $\mathbf{n} = (n_{1},\ \dots,\ n_{L})$.

Without loss of generality, let us consider a spin-$1/2$ as an example. A spin operator $\hat{\mathbf{S}}_l = (\hat{S}^x_l$, $\hat{S}^y_l$, $\hat{S}^z_l)$ defined at site-$l$ has three components in the $x$, $y$ and $z$-axes, respectively. For any state, there are only two free real variables out of two complex coefficients in front of the basis $\rvert \sigma_l \rangle$ of the $\hat{S}^z_l$-representation. The index $\sigma_l = \uparrow$, $\downarrow$, corresponding to values of $\pm 1/2$. These variables are governed by a pair of site-dependent angles $\theta_l$ and $\phi_l$ appearing in a spin-coherent state~\cite{Penc2010}
\begin{eqnarray}
\rvert \Omega_l \rangle = c^\uparrow_l \left\rvert \uparrow \right\rangle + c^\downarrow_l \left\rvert \downarrow \right\rangle\ ,\label{scs}
\end{eqnarray}
where the coefficients are given by
\begin{eqnarray}
c^\uparrow_l = \cos (\theta_l/2) \quad \text{and} \quad c^\downarrow_l = \sin (\theta_l/2) e^{ i\phi_l}\ .\label{coherentstate}
\end{eqnarray}
As convention, $\theta_l\in [0$, $\pi]$ and $\phi_l\in [0$, $2\pi)$. In such a state, $\mathbf{S}_l = \langle \hat{\mathbf{S}}_l \rangle = \mathbf{\Omega}_l / 2$ behaves as half of the unit vector in three-dimensional coordinates, that is,
\begin{eqnarray}
\mathbf{\Omega}_l=(\sin\theta_l \cos\phi_l,\ \sin\theta_l \sin\phi_l,\ \cos\theta_l)\ .
\end{eqnarray}
The phase factor for the basis $\rvert \sigma_l \rangle$ with a non-vanishing amplitude only depends on $\phi_l$, since both $\cos (\theta_l/2)$ and $\sin (\theta_l/2)$ are positive. Besides, a phase angle $h_l$ can modulate the phase factor in front of the spin-coherent state (\refeq{scs}), i.e., $e^{ih_l} \rvert \Omega_l \rangle$.

In the GWMF theory~\cite{Gutzwiller1963, Gutzwiller1965}, the wave function of the ground state $\rvert \psi^\text{gw} \rangle$ is a product of bases for $L$ spin-$1/2$s, i.e.,
\begin{eqnarray}
\rvert \psi^\text{gw} \rangle = \bigotimes^L_{l=1} \rvert \Omega_l \rangle = \sum_{\{\bm{\sigma}\}} \bar{a}^\text{gw}_{\bm{\sigma}} p^\text{gw}_{\bm{\sigma}} \rvert \bm{\sigma} \rangle\ ,\label{GWMF}
\end{eqnarray}
where
\begin{eqnarray}
\bar{a}^\text{gw}_{\bm{\sigma}} = \prod_l \lvert c^{\sigma_l}_l \rvert \quad \text{and} \quad p^\text{gw}_{\bm{\sigma}} = \exp\left[i \left(\bm{\phi} \cdot \mathbf{n} + \sum_{l=1}^{L} h_l\right)\right] \quad
\end{eqnarray}
represent the positive amplitude and the phase factor, respectively. Here, we define the angle vector as $\bm{\phi}=(\phi_1,\ \cdots,\ \phi_L)$, and the index vector $\bm{\sigma} = (\sigma_1,\ \cdots,\ \sigma_L)$ satisfies the corresponding relation $\mathbf{n} = 1/2 - \bm{\sigma}$. For a quantum model featuring a real Hamiltonian discussed in this work, the complex conjugate of the GWMF wave function $\rvert \psi^\text{gw} \rangle^*$ also indicates a state sharing the ground-state energy. Thus, the modified GWMF wave function $e^{i h_0} \vert \psi^\text{gw} \rangle = \vert \psi^\text{gw}_r \rangle + i \vert \psi^\text{gw}_i \rangle$ for the ground state can be decomposed into the real part $\vert \psi^\text{gw}_r \rangle$ and imaginary part $\vert \psi^\text{gw}_i \rangle$, given a specific global phase angle $h_0$. Both parts are real-valued and have an extra orthogonalization relation of $\langle \psi^\text{gw}_r \vert \psi^\text{gw}_i \rangle = \langle \psi^\text{gw}_i \vert \psi^\text{gw}_r \rangle = 0$. Concretely, two parts are expressed as
\begin{eqnarray}
\begin{split}
\rvert \psi^\text{gw}_r \rangle &= \sum_{\{\bm{\sigma}\}} \bar{a}^\text{gw}_{\bm{\sigma}} \cos(\bm{\phi} \cdot \mathbf{n} + \tilde{h}) \rvert \bm{\sigma} \rangle\ ,\\
\rvert \psi^\text{gw}_i \rangle &= \sum_{\{\bm{\sigma}\}} \bar{a}^\text{gw}_{\bm{\sigma}} \sin(\bm{\phi} \cdot \mathbf{n} + \tilde{h}) \rvert \bm{\sigma} \rangle\ ,
\end{split}
\end{eqnarray}
where the phase angle $\tilde{h} = \sum^L_{l=1} h_l + h_0$ is given. Regardless of whether the imaginary part $\vert \psi^\text{gw}_i \rangle$ is null or two parts share the same energy, we can always obtain a real ground-state wave function. It is worth noticing that the mean fields in the GWMF theory prefer selecting one of the degenerate manifolds if they exist, which artificially breaks the corresponding symmetry. So, the above-mentioned rotation, adjusted by the phase angle $h_0$, would introduce an energy split between the real part $\vert \psi^\text{gw}_r \rangle$ and the imaginary part $\vert \psi^\text{gw}_i \rangle$. In our theory, we ignore this effect and suppose their degeneracy still survives.

To assume that the real part $\rvert \psi^\text{gw}_r \rangle = \sum_{\{\bm{\sigma}\}} c^\text{gw}_{\bm{\sigma}} \rvert \bm{\sigma} \rangle$ can be expanded in the representation of bases $\rvert \bm{\sigma} \rangle$, the real expansion coefficient $c^\text{gw}_{\bm{\sigma}} = \bar{a}^\text{gw}_{\bm{\sigma}} \cos(\bm{\phi} \cdot \mathbf{n} + \tilde{h}) = s^\text{gw}_{\bm{\sigma}} a^\text{gw}_{\bm{\sigma}}$ comprises of an amplitude $a^\text{gw}_{\bm{\sigma}} = \lvert c^\text{gw}_{\bm{\sigma}} \rvert$ and a sign $s^\text{gw}_{\bm{\sigma}} = \text{Sgn} (c^\text{gw}_{\bm{\sigma}})$ following a rule:
\begin{eqnarray}
s^\text{gw}_{\bm{\sigma}} = \text{Sgn}[\cos(\bm{\phi} \cdot \mathbf{n} + \tilde{h})]\ .\label{LOSR}
\end{eqnarray}
The rule is called the leading-order version, removing short-range fluctuations completely. The phase angle $\tilde{h}$ is determined by other necessarily preserved global symmetries for a specified eigenstate, e.g., translational and inversion symmetries. And ``$\text{Sgn}$" denotes the standard sign function. When examining the sign rule for the nonzero imaginary part $\rvert\psi^\text{gw}_i \rangle$, an extra $\pi/2$ needs to be added to the phase angle $\tilde{h}$, which equivalently replaces the cosine function with the sine function in Eq.~(\ref{LOSR}). In the alternative notation-$\mathbf{n}$ of utilizing bases $\rvert \mathbf{n} \rangle$, the sign rule remains the same, i.e., $s^\text{gw}_{\mathbf{n}} \equiv s^\text{gw}_{\bm{\sigma}}$. For the convenience of the following discussions, we only use the notation-$\mathbf{n}$.

In the GWMF scenario, spins in the ordered state are typically visualized as classical vectors that follow a regular profile $\{\mathbf{\Omega}_l\}$ in space. Our derivation shows that the leading-order sign rule, which depends on angles $\phi_l$, is closely related to the spin-order profile $\{\mathbf{\Omega}_l\}$. The above conclusion is still valid for general qubit lattice models. In Sec.~\ref{secB}, we will demonstrate that the leading-order sign rule can, in principle, be learned by shn-FNN.

\section{Single-hidden-neuron feed-forward neural network}\label{secB}
The feed-forward neural network (FNN) is a powerful tool for approximating continuous functions~\cite{Cybenko1989, Hornik1989} and sorting samples by discrete values of characters~\cite{Goodfellow2016}. For instance, it can be applied to the classification of the double-valued sign $s_{\mathbf{n}}$ for arbitrary basis $\rvert \mathbf{n} \rangle$, when the expansion coefficient $c_{\mathbf{n}}$ in the ground-state wave function $\rvert \psi \rangle = \sum_{\mathbf{n}} c_{\mathbf{n}} \rvert \mathbf{n} \rangle$  for a real Hamiltonian consists of an amplitude $a_{\mathbf{n}}$ and a sign $s_{\mathbf{n}}$. However, as FNN becomes deeper, its complexity grows, making it difficult to understand the sign rule and its connection to meaningful physics insights. To address this issue, we introduce shn-FNN, similar to previous shallow FNNs~\cite{He2020, Yuan2021} but distinct from recently developed operations in a compact latent space~\cite{Iten2020, Wang2021}.

\begin{figure}[h]
\begin{center}
\includegraphics[width=\linewidth]{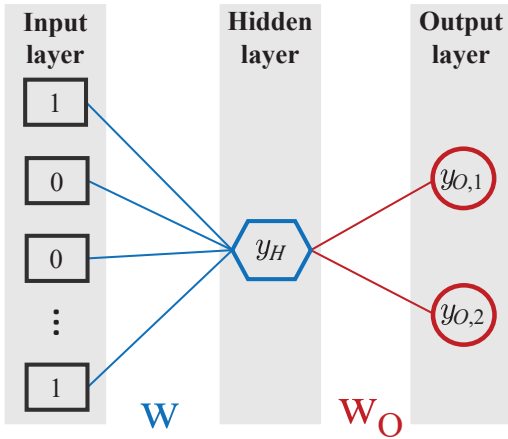}
\caption{Shn-FNN is utilized to acquire the leading-order sign rule of a quantum state for $L$ qubits. The network consists of three layers: the input layer (black squares) with $L$ neurons, the hidden layer (blue hexagons) with $1$ neuron, and the output layer (red circles) with $2$ neurons. These layers are connected by two weight vectors $\bm{w}$ (blue lines) and $\bm{w}_{O}$ (red lines). The hidden and output layers are activated using a cosine function and a $\text{softmax}$ function, respectively. The sign determined by shn-FNN is positive if $y_{O,1} > y_{O,2}$ and negative otherwise.
}\label{fig1}
 \end{center}
\end{figure}

The shn-FNN consists of an input, hidden, and output layer, as illustrated in Fig.~\ref{fig1}. The configuration $\mathbf{n}$ is assigned to the input layer of shn-FNN by simply setting a $L$-dimensional vector $\bm{y}_I = \mathbf{n}$. The hidden layer contains one neuron that produces a one-dimensional vector $y_{H}$. The output layer consists of two neurons that form a one-hot vector $\bm{y}_{O} = (y_{O,1},\ y_{O,2})$. These three layers are connected by two weight vectors $\bm{w}$ and $\bm{w}_{O}$.

The activation function cosine is empirically chosen for the hidden layer so that
\begin{eqnarray}
y_H = \cos(\bm{w} \cdot \mathbf{n})\ .
\end{eqnarray}
The vector $\bm{y}_{O}$ in the output layer is determined by applying the softmax function, i.e.,
\begin{eqnarray}
\bm{y}_{O} = \text{softmax} (y_{H} \bm{w}_{O})\ ,
\end{eqnarray}
where the weight vector $\bm{w}_{O} = (w_{O,1}$, $w_{O,2})$ is fixed. The function $\text{softmax}$ executes normalization by an exponential function to obtain probabilities, which is usually used in classification tasks~\cite{Goodfellow2016}. Specifically, the function $\text{softmax}$ gives two neurons $y_{O,1}$ and $y_{O,2}$ in the output layer, given by
\begin{eqnarray}
\begin{split}
y_{O,1} &= \frac{e^{w_{O,1} y_{H}}}{e^{w_{O,1} y_{H}} + e^{w_{O,2} y_{H}}}\ ,\\
y_{O,2} &= \frac{e^{w_{O,2} y_{H}}}{e^{w_{O,1} y_{H}} + e^{w_{O,2} y_{H}}}\ .
\end{split}
\end{eqnarray}
In this work, we choose $\bm{w}_{O} = (1, -1)$, so that $e^{w_{O,1} y_{H}}$ and $e^{w_{O,2} y_{H}}$ are distinguishable. Thus, the desired sign $s^\text{fnn}_\mathbf{n}$ is determined as follows:
\begin{eqnarray}
s^\text{fnn}_\mathbf{n} =
\left\{
\begin{array}{lc}
+1 & \quad \text{for} \quad y_{O,1} > y_{O,2}\\\\
-1 & \quad \text{for} \quad y_{O,1} \le y_{O,2}
\end{array}
\right.\ ,
\end{eqnarray}
which equivalently indicates the sign of $y_{H}$, i.e., $s^\text{fnn}_\mathbf{n} \equiv \text{Sgn}(y_H)$ in principle. Without one-hot representation, it has been proven that FNN performs worse since a categorization task turns into a regression task compulsively~\cite{Goodfellow2016}.

In summary, the shn-FNN representation for the sign rule corresponds to a function
\begin{eqnarray}
s^\text{fnn}_\mathbf{n} = \text{Sgn}\left[\cos(\bm{w} \cdot \mathbf{n})\right]\ .\label{eq1}
\end{eqnarray}
Compared to the leading-order sign rule~(\ref{LOSR}) that is derived from the GWMF theory,  we usually get
\begin{eqnarray}
w_l = \phi_l + \tilde{h} / N \quad \text{and} \quad N = \sum^L_{l=1} n_l\label{srconvert}
\end{eqnarray}
for models with constant $N$. Thus, Eq.~(\ref{eq1}) is also called the leading-order sign rule.

\section{Data sets and training}\label{secC}
{\it Data sets}. We use the exact diagonalization (ED) method to obtain the ground-state wave function. In the wave function, the sign $s_\mathbf{n}$ and the configuration $\mathbf{n}$ for a specified basis $\rvert \mathbf{n} \rangle$ constitute a sample in the data set $\mathbf{T}$. Each sign $s_\mathbf{n}$ is encoded as a one-hot vector $\bm{y}^{(\mathbf{n})} = (y^{(\mathbf{n})}_1$, $y^{(\mathbf{n})}_2)$, which can only take two valid combinations:
\begin{eqnarray}
\bm{y}^{(\mathbf{n})} =
\left\{
\begin{array}{lc}
(1,\ 0) & \quad \text{for} \quad s_{\mathbf{n}} > 0\\\\
(0,\ 1) & \quad \text{for} \quad s_{\mathbf{n}} < 0
\end{array}
\right.\ .
\end{eqnarray}

After arranging the samples in descending order of amplitude $a_\mathbf{n}$, we discard those with $a_{\mathbf{n}} < 10^{-15}$ to avoid any artificial effects caused by the limited numeric precision. The remaining $\mathcal{N}_s$ samples in the data set $\mathbf{T}$ are divided into a training set $\mathbf{T}_\text{train}$ and a testing set $\mathbf{T}_\text{test}$ in a $4:1$ ratio~\cite{Goodfellow2016}. Thus, the number of samples in the testing set $\mathbf{T}_\text{test}$ is given by $\mathcal{N}_\text{test} = \mathcal{N}_s / 5$.

{\it Training}. During the training scheme-$\rm{I}$, we employ the back-propagation (BP) algorithm~\cite{Rumelhart1986LIR} to optimize the variables in the weight vector $\bm{w}$ while adaptively adjusting the learning rate using the Adam algorithm~\cite{Adam2015}. The process aims to minimize the cross entropy, defined as
\begin{eqnarray}
{\cal S}_{\times}= -\sum_{\{\mathbf{n}\}} \left(y^{(\mathbf{n})}_{1}\ln y^{(\mathbf{n})}_{O,1} + y^{(\mathbf{n})}_{2}\ln y^{(\mathbf{n})}_{O,2}\right)\ ,\label{eq4}
\end{eqnarray}
which sums over samples in the entire training set. Here, the one-hot vector $\bm{y}^{(\mathbf{n})}_O = (y^{(\mathbf{n})}_{O,1}$, $y^{(\mathbf{n})}_{O,2})$ is the output of shn-FNN as the vector $\mathbf{n}$ is input.

We employ the mini-batch method based on the stochastic gradient descent (SGD)~\cite{Goodfellow2016, Wilson2003} to reduce the huge computational costs. Instead of using the entire training set $\mathbf{T}_\text{train}$ directly, we randomly select $\mathcal{N}_\text{step} = 100$ samples to calculate the gradients of the weights $w_l$ at each training step. In such a case, Eq.~(\ref{eq4}) only sums over selected $\mathcal{N}_\text{step}$ samples. This method performs well in accuracy and speed, and the random selection helps prevent the ``over-fitting" problem. In our program, we implement shn-FNN and the Adam optimization using the ML library ``TensorFlow"~\cite{Abadi2016}.

To evaluate the performance of shn-FNN, we introduce the accuracy rate ($\text{AR}$), which is calculated as the ratio of the number of successfully classified samples $\mathcal{N}^c_s$ to the total number of samples $\mathcal{N}_s$ in the data set $\mathbf{T}$, i.e., $\text{AR}=\mathcal{N}^c_s / \mathcal{N}_s$. To monitor the optimization process, we define two additional accuracy rates. At each training step, we calculate the number of samples successfully classified $\mathcal{N}^c_\text{step}$. Then, we evaluate the optimization by computing the accuracy rate $\text{AR}_\text{step} = \mathcal{N}^c_\text{step} / \mathcal{N}_\text{step}$. To provide a comprehensive evaluation, we utilize the testing set $\mathbf{T}_\text{test}$ and define the accuracy rate $\text{AR}_\text{test} = \mathcal{N}^c_\text{test} / \mathcal{N}_\text{test}$, where $\mathcal{N}^c_\text{test}$ represents the number of correctly classified samples in the testing set $\mathbf{T}_\text{test}$.

After each training step, we assess the convergence criterion $\epsilon$, which measures the absolute value of the difference between the accuracy rates $\text{AR}_\text{test}$ obtained in the current step and the previous step. We halt the training process once the convergent criterion $\epsilon$ falls below a threshold $\epsilon_0 = 10^{-3}$.

\begin{figure}[t]
\begin{center}
\includegraphics[width=\linewidth]{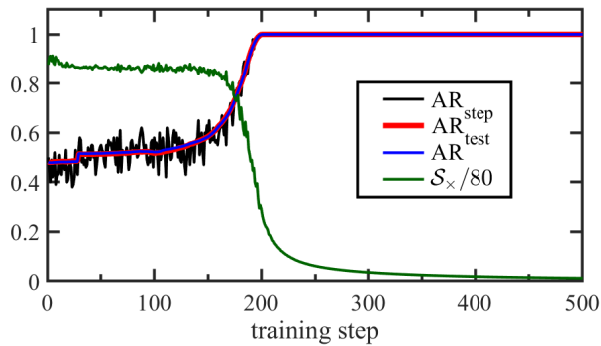}
\caption{Following scheme-$\rm{I}$, we train shn-FNN to acquire the leading-order sign rule (\refeq{eq1}) of the ground-state wave function for a generalized ferromagnetic ($J < 0$) Ising model (\refeq{HamGI}) of $L=16$ spins in a ring. The spin polarization is along the x-axis, i.e., $\mathbf{\Omega}=\hat{x}$. Throughout the training process, we monitor the cross entropy $\mathcal{S}_\times$ (green) as well as three different accuracy rates: $\text{AR}_\text{step}$ (black), $\text{AR}_\text{test}$ (red) and $\text{AR}$ (blue).
}\label{fig2}
\end{center}
\end{figure}

To exemplify how shn-FNN learns the sign rule, we study the ground state of a generalized Ising ring described by the Hamiltonian
\begin{eqnarray}
\hat{H}_{\text{Ising}} (J,\ \mathbf{\Omega}) = J \sum^{L}_{l=1} (\hat{\mathbf{S}}_l \cdot \mathbf{\Omega}) (\hat{\mathbf{S}}_{l+1} \cdot \mathbf{\Omega})\ ,\label{HamGI}
\end{eqnarray}
where $J$ represents the strength of the coupling, and $\mathbf{\Omega}$ determines the orientation of the spin polarization.

For the case of ferromagnetic coupling ($J < 0$) and spin polarization along the x-axis ($\mathbf{\Omega} = \hat{x}$), we obtain the ground-state wave function for $L=16$ spins, and then prepare the data sets $\mathbf{T}$, $\mathbf{T}_\text{train}$ and $\mathbf{T}_\text{test}$, as stated above. We initialize the weight vectors $\bm{w}$ and $\bm{w}_O$ in shn-FNN and start the training process. As illustrated in Fig.~\ref{fig2}, the cross entropy $\mathcal{S}_\times$ rapidly decreases around the $170$-th training step. After approximately $200$ training steps, the cross entropy $\mathcal{S}_\times$ tends towards stability, and the three accuracy rates $\text{AR}_\text{step}$, $\text{AR}_\text{train}$ and $\text{AR}$ consistently reach a maximum value $1$ or $100\%$. We terminate the training process when the convergence criterion $\epsilon < \epsilon_0$ is met. In addition to scheme-$\rm{I}$, we will introduce scheme-$\rm{II}$ in Sec.~\ref{subsecB1}, tailored for frustrated spin models and the Fermi-Hubbard ring later.

\section{Qubit lattice models}\label{secD}
Using shn-FNN, we analyze the leading-order sign rules for various ordered ground states in qubit lattice models, including non-frustrated spin models in Sec.~\ref{subsecA}, frustrated spin models in Sec.~\ref{subsecB}, and interacting fermions in Sec.~\ref{subsecC}.

\subsection{Non-frustrated spin models}\label{subsecA}

\begin{figure}[t]
\begin{center}
\includegraphics[width=\linewidth]{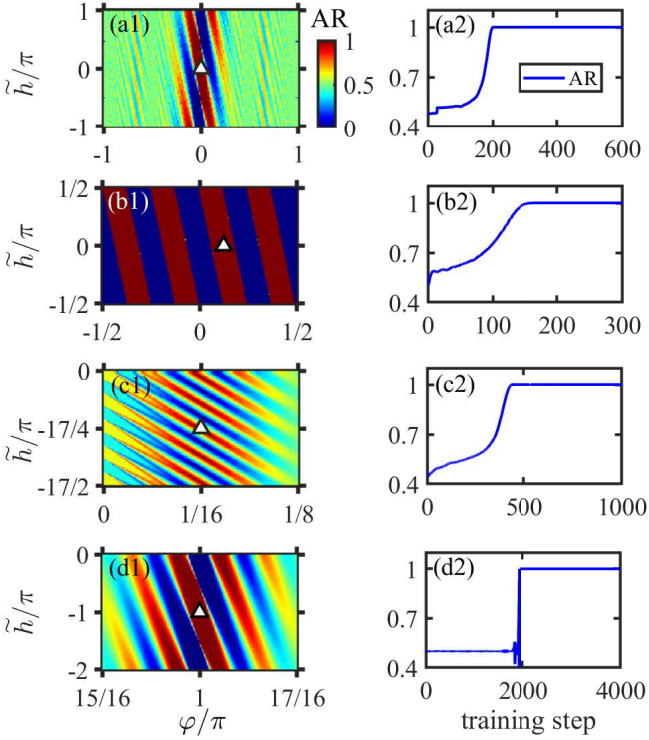}
\caption{(a.1-d.1) The accuracy rate $\text{AR}$ of the sign rule~(\refeq{eq1}) as a function of the angle $\varphi$ and phase angle $\tilde{h}$. In (a.1) and (b.1), the angle $\phi_l = \varphi$, while in (c.1) and (d.1) $\phi_l = \varphi l$. (a.2-d.2) Following scheme-$\rm I$, the accuracy rate $\text{AR}$ changes with the training step number during the training process. After full training, the optimized shn-FNN with $\text{AR}=1$ gives weights $w_l$. These weights are then converted to $\phi_l$ and $\tilde{h}$ using Eq.~(\refeq{srconvert}), which follow simple rules: $\phi_l = \varphi$ in (a.2) and (b.2), while $\phi_l = \varphi l$ in (c.2) and (d.2). Consequently, we mark coordinates $\varphi$ and $\tilde{h}$ in (a.1-d.1) by black open triangles for representing the optimized weights. Here, we investigate ground states in (a.1, a.2) a generalized ferromagnetic Ising ring with $\mathbf{\Omega}=\hat{x}$, (b.1, b,2) a ferromagnetic spin-$1/2$ XY ring, (c.1, c.2) a twisted ferromagnetic spin-$1/2$ XY ring (even parity), and (d.1, d.2) an antiferromagnetic spin-$1/2$ XY ring. The system size is $L=16$.
}\label{fig3}
\end{center}
\end{figure}

\subsubsection{A generalized ferromagnetic Ising ring}\label{subsecA1}
For the case of $J < 0$ and $\mathbf{\Omega} = \hat{x}$ in the model (\ref{HamGI}), the optimized shn-FNN with $\text{AR}=1$, as illustrated in Fig.~\ref{fig3}(a.2), suggests that the weights $w_{l}=0$ in the sign rule~(\ref{eq1}). The physical interpretation of $w_l$ can be understood from the connection between Eq. (\ref{eq1}) and the leading-order sign rule~(\ref{LOSR}). Using Eq.~(\ref{srconvert}), $w_l = 0$ can be converted to a combination of the angles $\phi_l=0$ and the phase angle $\tilde{h}=0$ in the sign rule~(\ref{LOSR}). This demonstrates the presence of ferromagnetic order along the x-axis, according to the spin-coherent state representation \eqref{coherentstate}. To assume that $\phi_l = \varphi$ in Eq.~(\ref{srconvert}), we can visualize AR for the sign rule~(\ref{eq1}) in the ($\varphi$, $\tilde{h}$) plane, as shown in Fig.~\ref{fig3}(a.1). The optimized weights $w_l=0$ are positioned at the coordinates of a maximum, i.e., $\varphi=0$ and $\tilde{h}=0$ (black open triangle).

\subsubsection{A ferromagnetic spin-$1/2$ XY ring}\label{subsecA2}
For a spin-$1/2$ XY ring with $L=16$ sites, the Hamiltonian is given by
\begin{eqnarray}
\hat{H}^P_{\text{xy}} (J) = \frac{1}{2} \left(\sum^L_{l=1} J \hat{S}^{+}_{l} \hat{S}^{-}_{l+1} + \textrm{h.c.}\right)\ ,
\end{eqnarray}
where $\hat{S}^{\pm}_l$ represent the flipping-up and flipping-down operators for a spin-$1/2$. The coupling strengths $J$ in the $x$ and $y$-axes are equal in this ring.

For the case of $J < 0$ shown in Fig.~\ref{fig3}(b.2), the accuracy rate $\text{AR}$ for shn-FNN reaches a perfect-classification limit of $1$ after approximately $150$ training steps. In the optimized shn-FNN, the  weights are given by $w_{l} = \pi / 8$, which are equivalent to $\phi_l=\pi/8$ and $\tilde{h}=0$ according to Eq.~(\ref{srconvert}). The result remains consistent with an in-plane ferromagnetic order, where all spins are confined to the xy-plane and aligned in the same polarization direction. Therefore, we still set $\phi_l=\varphi$ in Eq.~(\ref{srconvert}), and plot the AR distribution in the ($\varphi$, $\tilde{h}$) plane in Fig.~\ref{fig3}(b.1). We can easily find that the weights $w_l=\pi/8$ correspond to the coordinates $\varphi = \pi / 8$ and $\tilde{h}=0$ (black open triangle). Since $w_l = \pi/8$ is uniform in space, the resulting sign rule (\refeq{eq1}) can be summarized as the Perron-Frobenius theorem~\cite{Perron1907, Frobenius1909} by removing a global phase angle $\pi$, when $L=16$.

\subsubsection{A twisted ferromagnetic spin-$1/2$ XY ring}\label{subsecA3}
For a spin-$1/2$ XY ring with the ferromagnetic coupling strength $J < 0$ under the twisted boundary condition (TBC), an antiferromagnetic bond connects site-$1$ and site-$L$ in the Hamiltonian
\begin{eqnarray}
\hat{H}^{T}_{\text{xy}} (J) = \frac{1}{2} \left(\sum^{L-1}_{l=1} J \hat{S}^+_l \hat{S}^-_{l+1} - J \hat{S}^+_L \hat{S}^-_1 + \textrm{h.c.}\right)\ .\quad
\end{eqnarray}
To achieve convergence of the accuracy rate $\text{AR} = 1$, we train the shn-FNN for $L=16$, as shown in Fig.~\ref{fig3}(c.2). The optimized weights obtained from this training are given by
\begin{eqnarray}
w_l = \frac{\pi}{L} l - \frac{(L+1)\pi}{2L} \label{WXY}
\end{eqnarray}
in the sign rule~(\ref{eq1}) for the ground-state wave function. We find Eq.~\eqref{WXY} corresponds to a combination of the angles $\phi_l = \pi l / 16$ and the phase angle $\tilde{h}=-17\pi / 4$ according to Eq.~(\ref{srconvert}). This suggests a gradually varying spin profile in space with a pitch angle of $\pi / 16$. In a similar manner, by setting $\phi_l = \varphi l$ in Eq.~(\ref{srconvert}), we parameterize the sign rule~(\ref{eq1}) with two parameters $\varphi$ and $\tilde{h}$. The AR distribution in the ($\varphi$, $\tilde{h}$) plane is plotted in Fig.~\ref{fig3}(c.1), where the coordinates $\varphi = \pi / 16$ and $\tilde{h}=-17\pi / 4$ are marked by a black open triangle.

The above result can be better understood through the following analysis. Under a rotation defined by the operators
\begin{eqnarray}
\hat{\mathcal{U}}_{\varphi} = \prod^L_{l=1} \hat{\mathcal{R}}_l (l \varphi) \quad \text{and} \quad \hat{\mathcal{R}}_l (q) = e^{i q \hat{S}^z_l}\ , \quad
\end{eqnarray}
the twisting effect from the antiferromagnetic bond is absorbed into a gauge field $\tilde{J}=\exp(i \varphi)$ in a new Hamiltonian ${\hat H}^{P}_{\text{xy}}(J\tilde{J})$, where $\varphi=\pi/L$. Meanwhile,
\begin{eqnarray}
\hat{\mathcal{U}}^{\dagger}_{\varphi} {\hat S}^{\pm}_{l} \hat{\mathcal{U}}_{\varphi} = {\hat S}^{\pm}_{l} \exp(\mp i l \varphi)\ .
\end{eqnarray}
Based on the arguement in App.~\ref{appC}, it is found that the even-parity ground-state wave function $\rvert \psi^{P}_{\text{xy}}(J\tilde{J}) \rangle$ for the Hamiltonian ${\hat H}^{P}_{\text{xy}}(J\tilde{J})$ has positive signs, i.e., $s_{\mathbf n} > 0$. As a result, the ground-state wave function $\rvert \psi^{T}_{\text{xy}}(J) \rangle$ for the Hamiltonian ${\hat H}^{T}_{\text{xy}}(J)$ carries a nonzero complex phase factor due to the rotation $\hat{\mathcal{U}}_\varphi$. Specifically, it can be written as
\begin{eqnarray}
\rvert \psi^{T}_{\text{xy}}(J) \rangle = \hat{\mathcal{U}}_{\varphi} \rvert \psi^{P}_{\text{xy}}(J\tilde{J}) \rangle\ .
\end{eqnarray}
Thus, the real part of the wave function is given by
\begin{eqnarray}
|\psi^{T}_{\text{xy},r}(J)\rangle= \cos(\bm{w} \cdot \mathbf{n}) \rvert \psi^{P}_{\text{xy}}(J\tilde{J}) \rangle\ ,\label{txysr}
\end{eqnarray}
which is inversion-symmetric concerning the chain center. It is worth noting that although Eq.~(\refeq{txysr}) implies the same sign rule (\refeq{eq1}) obtained from the GWMF theory, this analysis is rigorous for the twisted spin-$1/2$ XY ring.

\subsubsection{An antiferromagnetic spin-$1/2$ XY ring}\label{subsecA4}
For the case of antiferromagnetic coupling $J>0$, the optimized shn-FNN suggests $w_l = \pi l - \pi/8$ in the sign rule~(\ref{eq1}), as shown in Fig.~\ref{fig3}(d.1). We find that $\bm{w}$ corresponds to a combination of the angles $ \phi_l = \pi l$ and the phase angle $\tilde{h}=-\pi$, indicating the presence of N$\acute{e}$el order. Remarkably, the sign rule defined by $\bm{w}$ is equivalent to MPR, where $s_{\mathbf n}=(-1)^{N_{O}}$, and the quantity
\begin{eqnarray}
N_{O} = \sum_{l\in \text{odd}} \langle {\hat S}^{z}_{l} + 1/2 \rangle
\end{eqnarray}
sums over all odd sites. Here, we parameterize the sign rule (\ref{eq1}) with two parameters $\varphi$ and $\tilde{h}$, by setting $\phi_l=\varphi l$ in Eq.~(\ref{srconvert}). Fig.~\ref{fig3}(d.1) illustrates the AR distribution in the ($\varphi$, $\tilde{h}$) plane, with the corresponding coordinates of the optimized $\bm{w}$ represented by a black open triangle, specifically $\varphi = \pi$ and $\tilde{h}=-\pi$. This sign rule is well understood because the ground-state wave function for $J>0$ is connected to the one for $J<0$ by a $\pi$-rotation operation
\begin{eqnarray}
\hat{\mathcal{U}}_{\pi} = \prod^L_{l=1} \hat{\mathcal{R}}_l (l \pi) = \prod_{l\in\text{odd}} \hat{\mathcal{R}}_l (\pi)\ .
\end{eqnarray}

Hence, it is evident that the sign rules in the XY ring with perfect AR, shown in Sec.~\ref{subsecA2}, \ref{subsecA3}, and \ref{subsecA4}, adhere to a standardized format of weights Eq.~(\ref{srconvert}) by setting $\phi_l = \varphi l$ with specific pitch angles $\varphi=0$, $\pi / L$ and $\pi$. This pitch angle $\varphi$ is related to the profile of spins rotating in space and can be acquired by training shn-FNN.

\subsubsection{An antiferromagnetic Heisenberg ring}\label{subsecA5}
In a pure antiferromagnetic Heisenberg ring (AFHR) with equal nearest-neighboring antiferromagnetic couplings $J_1$ in the $x$, $y$, and $z$-axes, the spins at odd sites align anti-parallel to the spins at even sites according to GWMF. Even though the precise ground state behaves as the Tomonago-Luttinger liquid (TLL)~\cite{Tomonaga1950, Luttinger1963, Haldane1981}, the optimized shn-FNN suggests MPR, which is consistent with previous studies~\cite{Marshall1955, Richte1994}. We discuss the sign rule uniformly in the $J_1$-$J_2$ AFHR in Sec.~\ref{subsecB1}.

\subsection{Frustrated spin models}\label{subsecB}

\begin{figure}[b]
\begin{center}
\includegraphics[width=\linewidth]{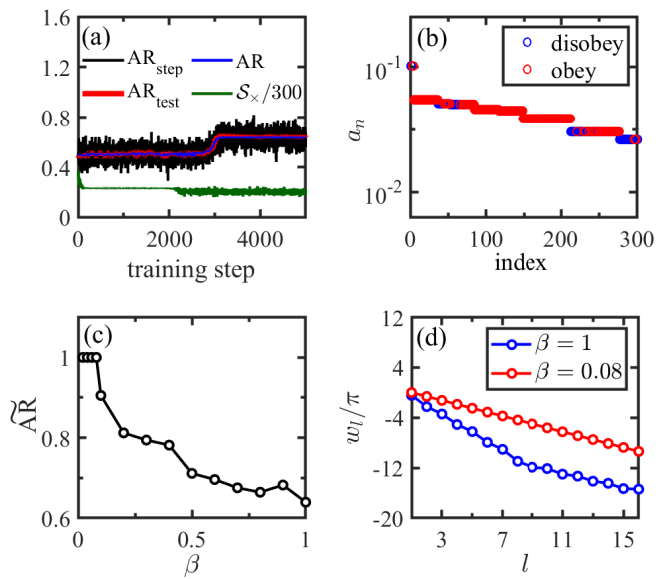}
\caption{Training shn-FNN to learn the leading-order sign rule (\refeq{eq1}) of the ground-state wave function for a $J_1$-$J_2$ antiferromagnetic Heisenberg model (\refeq{HamJ1J2}) of $L=16$ spins in a ring. The chosen ratio is $\alpha = 0.9$. (a) In scheme-$\rm{I}$, all $\mathcal{N}_s$ samples in the data set $\mathbf{T}$ are utilized for training shn-FNN. The training process is monitored by the cross entropy $\mathcal{S}_\times$ (green) as well as three distinct accuracy rates: $\text{AR}_\text{step}$ (black), $\text{AR}_\text{test}$ (red) and $\text{AR}$ (blue). (b) The samples are sorted in descending order of magnitude $a_\mathbf{n}$. They are then regrouped into two data sets: $\mathbf{T}_\text{obey}$ that follows the sign rule proposed by the optimized shn-FNN obtained from training scheme-$\rm{I}$ (red), and $\mathbf{T}_\text{disobey}$ that violates the sign rule (blue). (c) In scheme-$\rm{II}$, the first $\widetilde{\mathcal{N}}_s \le \mathcal{N}_s$ samples in the data set $\mathbf{T}$ are used for training shn-FNN. The accuracy rate $\widetilde{\text{AR}}$ is plotted as a function of the selection rate $\beta = \widetilde{\mathcal{N}}_s / \mathcal{N}_s$. (d) The spatial distribution of the weight vector elements $w_l$ in the sign rule (\refeq{eq1}) from training scheme-$\rm{I}$ (blue) and scheme-$\rm{II}$ (red) are shown.
}\label{fig4}
\end{center}
\end{figure}

\subsubsection{A $J_1$-$J_2$ antiferromagnetic Heisenberg ring}\label{subsecB1}
When the antiferromagnetic next-nearest-neighboring (NNN) Heisenberg coupling $J_2 > 0$ is introduced, we investigate the behavior of the frustrated spin-$1/2$ $J_1$-$J_2$ AFHR. The Hamiltonian for this system is given by
\begin{eqnarray}
\hat{H}_{J_1\text{-}J_2} = \sum_{l=1}^L ( J_1 \hat{\mathbf{S}}_{l} \cdot \hat{\mathbf{S}}_{l+1} + J_2 \hat{\mathbf{S}}_{l} \cdot \hat{\mathbf{S}}_{l+2})\ ,\label{HamJ1J2}
\end{eqnarray}
where $\alpha = J_2 / J_1$ is a dimensionless ratio.

Using the techniques suggested in Sec.~\ref{secC} and following training scheme-$\rm{I}$ exemplified in Fig.~\ref{fig2}, we initially train shn-FNN with all $\mathcal{N}_s$ samples in the data set $\mathbf{T}$. For example, when the ratio $\alpha=0.9$, the optimization of shn-FNN searches for the minima of the cross entropy $\mathcal{S}_\times$ with extremely low efficiency. As illustrated in Fig.~\ref{fig4}(a), three accuracy rates oscillate irregularly near the worst-performance limit of $0.5$. After approximately $3000$ training steps, all accuracy rates suddenly increase and reach another plateau. During this phase, the cross entropy $\mathcal{S}_\times$ exhibits random oscillation and fails to offer a meaningful gradient direction for updating the weight vector $\bm{w}$ in shn-FNN. Once the accurate rate $\text{AR}$ reaches approximately $0.63$, the weight vector $\bm{w}$ in the optimized shn-FNN (blue circles), as shown in Fig.~\ref{fig4}(d), becomes difficult to interpret.

To address this issue, we sort samples in the descending order of amplitude $a_\mathbf{n}$, as shown in Fig.~\ref{fig4}(b), and we observe that correct classification in the data set $\mathbf{T}_\text{obey}$ and wrong classification in the data set $\mathbf{T}_\text{disobey}$ are irregularly mixed. Instead, we adopt training scheme-$\rm{II}$, where we use the first $\widetilde{\mathcal{N}}_s \le \mathcal{N}_s$ samples in the data set $\mathbf{T}$ to train shn-FNN. In Fig.~\ref{fig4}(c), as we reduce the selection rate $\beta = \widetilde{\mathcal{N}}_s / \mathcal{N}_s$, the accuracy rate $\widetilde{\text{AR}} = \widetilde{\mathcal{N}}^c_s / \widetilde{\mathcal{N}}_s$ approaches the perfect-classification limit of $1$. With the optimized shn-FNN, $\widetilde{\mathcal{N}}^c_s$ out of $\widetilde{\mathcal{N}}_s$ samples are correctly classified. The resulting weight vector $\bm{w}$, shown in Fig.~\ref{fig4}(d), exhibits a straight line in the sign rule (\refeq{eq1}), which will be used to demonstrate physical insight later.

\begin{figure}[t]
\begin{center}
\includegraphics[width=\linewidth]{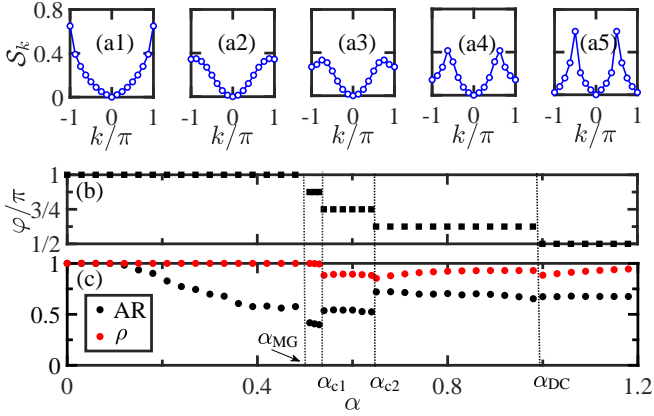}
\caption{(a.1-a.5) The structure factor $\mathcal{S}_k$ as a function of the momentum $k$ for a $J_1$-$J_2$ antiferromagnetic Heisenberg ring (\refeq{HamJ1J2}) with $L=16$. Five different values of the ratio $\alpha = J_2 / J_1$ are considered: (a.1) $0.35$, (a.2) $0.52$, (a.3) $0.58$, (a.4) $0.73$, and (a.5) $1.2$. (b) The optimized shn-FNN is trained using scheme-{\rm{II}}, and the resulting weights Eq.~(\ref{srconvert}) with $\phi_l = \varphi l$ gives the pitch angle $\varphi$. (c) Meanwhile, the corresponding accuracy rate $\text{AR}$ (black) and the correct weight $\rho$ (red) are also shown. The plots demonstrate a series of level crossings at $\alpha_{\text{MG}}=0.5$, $\alpha_{c1}\approx 0.53$, $\alpha_{c2} \approx 0.64$, and $\alpha_{\text{DC}} \approx 0.99$.
}\label{fig5}
\end{center}
\end{figure}

After conducting a systematical analysis of the ground states for $L=16$ sites, we have discovered that the optimized shn-FNN proposes the sign rules with weights Eq.~(\ref{srconvert}) by setting $\phi_l = \varphi l$, where the phase angle $\varphi = 2 p \pi / L$, as shown in Fig.~\ref{fig5}(b). The integer number $p$ ranges from $L / 2$ to $L / 4$, which allows us to divide the broad regime $\alpha \in [0,\ 1.2]$ into five intervals. Within each interval, the value of $\varphi$ plays the role of the commensurate/incommensurate pitch angle~\cite{White1996, Soos2016}. In Fig.~\ref{fig5}(a.1-a.5), we observe that the double peaks in the structure factor
\begin{eqnarray}
{\cal S}_{k} = \frac{1}{L^2} \sum_{l,l'} e^{i k (l - l')} \langle \hat{\mathbf{S}}_l \cdot \hat{\mathbf{S}}_{l'} \rangle
\end{eqnarray}
are located at the momenta $k= \pm \varphi$. Due to the interplay of interactions, the accuracy rate $\text{AR} < 1$ and the correct weight
\begin{eqnarray}
\rho = \sum_{\mathbf{n} \in \mathbf{T}_{\text{obey}}} |a_{\mathbf{n}}|^{2} < 1
\end{eqnarray}
are shown in Fig.~\ref{fig5}(c), where the data set $\mathbf{T}_{\text{obey}}$ includes samples obeying the leading-order sign rule (\ref{eq1}). Besides, we find that the correct weight $\rho$ in the whole parameter regime is close to $1$, which means that most of the samples with the largest amplitudes obey the leading-order sign rule, so the proposed scheme-$\rm{II}$ works well.

The investigation of the sign rule provides consistent physical insights. At the Majumdar-Ghosh (MG) point $\alpha_{\text{MG}}=0.5$, and for $\varphi=\pi$ within the range $\alpha \in [0,\ \alpha_{\text{MG}}]$, the leading-order sign rule is in accordance with MPR. As the ratio $\alpha$ approaches infinity, one of the decoupled chains, composed of odd or even sites, individually follows MPR. However, away from that limit, a relatively tiny positive $J_1$ promotes a stable commensurate spin order with a pitch angle $\varphi = \pi/2$. When the ratio $\alpha < \alpha_{\text{DC}} \approx 0.99$, commensurability is disrupted due to the emergence of triplet defects~\cite{White1996, Soos2016}. Between $\alpha_{\text{MG}}$ and $\alpha_{\text{DC}}$, the ground state undergoes an incommensurate crossover~\cite{White1996, Soos2016}, which is indicated by the varying pitch angle $\varphi$ in the weights of the leading-order sign rule (\refeq{eq1}), as shown in Fig.~\ref{fig5}(b).

Besides, the ground state maintains the translation symmetry with a conserved momentum of either $0$ or $\pi$, depending on the integer number $p$, so that $\phi_l = \varphi l$ in the leading-order sign rule~(\ref{LOSR}). Moreover, the center inversion symmetry of the chain imposes a constraint of $\tilde{h} = p \pi/2$. Consequently, in the equivalent sign rule~(\ref{eq1}), we choose the activation function sine/cosine for odd/even values of $p$.

\begin{figure}[t]
\begin{center}
\includegraphics[width=\linewidth]{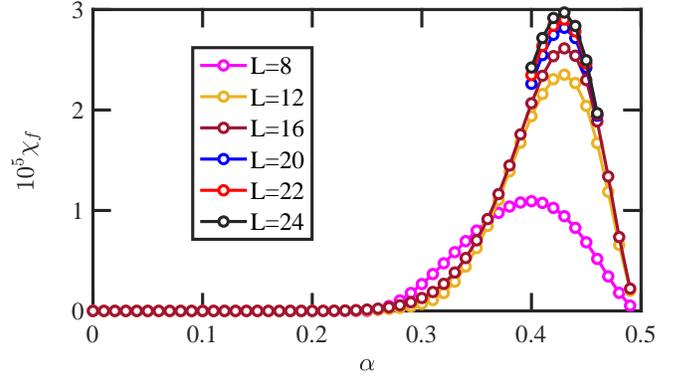}
\caption{The sign-fidelity susceptibility density $\chi_f$ for the ground state is examined in the region $\alpha\in [0$, $\alpha_\text{MG})$ for different system sizes $L=8$, $12$, $16$, $20$, $22$ and $24$ in the $J_1$-$J_2$ antiferromagnetic Heisenberg ring (\refeq{HamJ1J2}).
}\label{fig6}
\end{center}
\end{figure}

To quantitatively assess the violation of MPR when $\alpha \in [0,\ \alpha_\text{MG})$, we introduce a sign-fidelity
\begin{eqnarray}
f = \langle \psi^{\text{MPR}} \vert \psi \rangle\ .
\end{eqnarray}
Here, we define the MPR state $\vert \psi^{\text{MPR}} \rangle$ as
\begin{eqnarray}
\rvert \psi^{\text{MPR}} \rangle = \sum_{\{\mathbf{n}\}} s^{\text{MPR}}_{\mathbf{n}} a_{\mathbf{n}} \rvert \mathbf{n} \rangle\ ,
\end{eqnarray}
where the sign $s^{\text{MPR}}_{\mathbf{n}}$ fully satisfies MPR. Thus, we get $f = 2\rho-1$~\cite{Retzla1993, Richte1994, Zeng1995}. In the vicinity of a continuous transition point, the minimum sign-fidelity $f$ or correct weight $\rho$  is expected to be achieved, indicating the most complicated sign rule~\cite{Cai2018, Westerhout2020}. Like the orthogonalization catastrophe for free fermions~\cite{Anderson1967}, fidelity follows a pow-law function of the system size $L$. In principle, the relevant sign-fidelity susceptibility density, given by
\begin{eqnarray}
\chi_{f} = -(\ln f) / L\ ,
\end{eqnarray}
is capable of identifying the places of continuous transition points~\cite{Gu2010}. However, the maximum of $\chi_{f}$ is located at $\alpha_{\text{peak}}\approx 0.43 > \alpha_{\text{BKT}} \approx 0.241$ in the dimerized (DM) region~\cite{Wang2009}, where $\chi_f$ approaches a $L$-independent function as shown in Fig.~\ref{fig6}. It is possibly caused by the anomalous behavior of the exponential closure of gaps at the famous Berzinskii-Kosterlitz-Thouless (BKT) transition point $\alpha_{\text{BKT}}$~\cite{Cincio2019}.

\subsubsection{A spin-$1/2$ triangular XY antiferromagnet on a torus}\label{subsecB2}
\begin{figure}[t]
\begin{center}
\includegraphics[width=\linewidth]{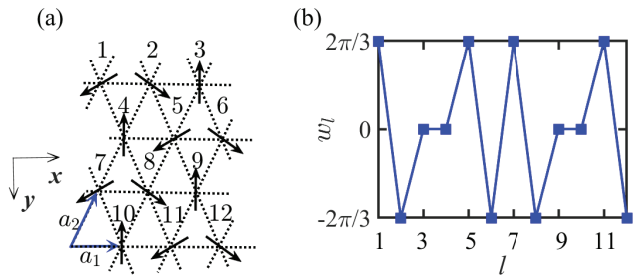}
\caption{(a) The spin-$1/2$ XY antiferromagnet (\refeq{HamTri}) with the XC geometry on a torus. $a_1$ and $a_2$ denote two primitive vectors. All sites are labeled with a single index $l$. (b) For the geometry $3\times4$, we employ scheme-$\rm{II}$ to train shn-FNN for the ground-state wave function. The spatial distribution of weights $w_l$ in the optimized shn-FNN is plotted.
}\label{fig7}
\end{center}
\end{figure}

Shn-FNN can learn the leading-order sign rules for the ground-state wave function of $2$D quantum models, such as the XY model on triangular lattices with a size of $L_x \times L_y$ sites, as shown in Fig.~\ref{fig7}(a). The corresponding Hamiltonian for the model reads
\begin{eqnarray}
{\hat H}_{\triangle} = \frac{1}{2} \sum_{\langle l,l'\rangle} (\hat{S}^+_l \hat{S}^-_{l'} + \textrm{h.c.})\ ,\label{HamTri}
\end{eqnarray}
and $\langle \rangle$ sums over all nearest-neighboring sites $l$ and $l'$. In the XC geometry, the lattice site labeled as $l = l_y L_x + l_x + 1$, is identified by binary indices $(l_x, l_y)$ with $l_x = 0$, $\cdots$, $L_x - 1$ and $l_y = 0$, $\cdots$, $L_y - 1$. The displacement for the site is given by $\mathbf{r}_l$.

\begin{figure}[b]
\begin{center}
\includegraphics[width=\linewidth]{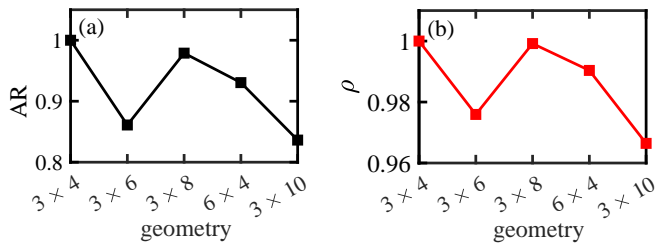}
\caption{Regarding the ground state of the spin-$1/2$ XY antiferromagnet on a torus (\refeq{HamTri}), we demonstrate (a) the accuracy rate $\text{AR}$ (black) and (b) correct weight $\rho$ (red) for the sign rule (\refeq{eq1}) with weights defined in Eq.~(\refeq{srTri}). Both are shown as a function of the geometry $L_x \times L_y$. It is worth noting that the optimized shn-FNN indicates that $\tilde{h} = 0$ for even $(L_y/2)$, but $\tilde{h} = \pi/2$ for odd $(L_y/2)$.
}\label{fig8}
\end{center}
\end{figure}

To ensure an exact hit at relevant high-symmetry momentum points $\mathbf{K}^{\pm}$ in the first Brillouin zone, the length $L_x$ is chosen as a multiple of $3$. Following training scheme-$\rm{II}$, the weights $w_{(l_x, l_y)}$ in the leading-order sign rule (\refeq{eq1}) are determined by the optimized shn-FNN. Specifically, we get
\begin{eqnarray}
w_{(l_x, l_y)} = \frac{2\pi}{3} (l_x + [l_y])\ ,\label{srTri}
\end{eqnarray}
where $[l_y]=1$ if $l_y$ is even and $0$ otherwise, as illustrated in Fig.~\ref{fig7}(b). This result matches the physical scenario of the coplanar $120^{\circ}$ order~\cite{Bach2021} observed in previous studies, where the angle between spin polarization orientations at neighboring sites is always $2\pi/3$.

Moreover, the ground state possesses point group symmetries of the torus. These symmetry operations listed in table~\ref{tab1} carry eigenvalues of $+1$ or $-1$, corresponding to the symmetric/even or antisymmetric/odd sector of the group representation in mathematics.

\begin{table}[ht]
\caption{The symmetry operations are measured on the ground-state wave function for the spin-$1/2$ triangular XY antiferromagnet on a torus. The table includes the translation by a site ${\cal T}_{x}$ in the x-axis and ${\cal T}_{y}$ in the y-axis, mirror inversion ${\cal M}_x$ about the x-axis and ${\cal M}_y$ about the y-axis, center inversion ${\cal I}_{c}$.}\label{tab1}
\centering
\begin{tabular}{ c  c  c  c  c  c }
\hline\hline
$L_{x} \times L_{y}$ & $\langle {\cal T}_{x} \rangle$ & $\langle {\cal T}_{y} \rangle$ & $\langle {\cal M}_{x} \rangle$ & $\langle {\cal M}_{y} \rangle$ & $\langle {\cal I}_{c} \rangle$ \\
\hline
$3 \times 4$ & $+1$ & $+1$ & $+1$ & $+1$ & $+1$ \\
$3 \times 6$ & $+1$ & $+1$ & $+1$ & $-1$ & $-1$ \\
$3 \times 8$ & $+1$ & $+1$ & $+1$ & $+1$ & $+1$ \\
$6 \times 4$ & $+1$ & $+1$ & $+1$ & $+1$ & $+1$ \\
$3 \times 10$ & $+1$ & $+1$ & $+1$ & $-1$ & $-1$ \\
\hline\hline
\end{tabular}%
\label{table:exponent}
\end{table}

Let us discuss the mirror inversion ${\cal M}_{y}$ about the y-axis. Under ${\cal M}_{y}$, the basis $|\mathbf{n}\rangle$ becomes $|\mathbf{n}'\rangle$ but the sign is unchanged, so we have
\begin{eqnarray}
\bm{w} \cdot \mathbf{n}' = L_{y} \pi - \bm{w} \cdot \mathbf{n} \quad \text{modulo} \quad 2 \pi\ .
\end{eqnarray}
For even $(L_{y}/2)$, such as the $3 \times 4$ lattice, the ${\cal M}_{y}$-symmetric ground-state wave function leads to $\tilde{h}=0$ in the leading-order sign rule~(\ref{LOSR}), or equivalently the activation function cosine in Eq.~(\ref{eq1}), which obeys
\begin{eqnarray}
\text{Sgn}[\cos(\bm{w} \cdot \mathbf{n})]=\text{Sgn}[\cos(\bm{w} \cdot \mathbf{n'})]\ .
\end{eqnarray}
In contrast, for odd $(L_{y}/2)$, e.g., the geometry of $3 \times 6$ lattices, the ${\cal M}_{y}$-antisymmetric ground-state wave function prefers $\tilde{h}=\pi/2$ in Eq.~(\ref{LOSR}) and the sine function in Eq.~(\ref{eq1}). This difference is captured by shn-FNN in Fig.~\ref{fig8}.

Furthermore, based on the mean-field picture of spinless Dirac fermions coupled to Chern-Simons gauge fields~\cite{Wang2018, Sedrakyan2020}, it has been shown that for different lattice geometries with finite $L_x$ and $L_y$, non-condensed BCS pairs of spinons from high symmetry points $\mathbf{K}^\pm$ would violate the leading-order sign rule, where both $\text{AR}$ and $\rho$  deviate from $1$. However, a more nuanced understanding of the subtle relationship between lattice geometry and the deviation from GWMF still needs to be included.

\subsection{A Fermi-Hubbard ring}\label{subsecC}
The Fermi-Hubbard model is a simple model that describes the physics in strongly correlated electron systems, which is closely connected to quantum magnetism, metal-insulator transition, and the promising theory of high-temperature superconductivity~\cite{Henderson1992, Essler2005, Moriya2012}. In a ring, the Hamiltonian for two-species fermions can be written as
\begin{eqnarray}
\hat{H}_\text{F}= \sum^{L}_{l=1} \left[-t\sum_{\sigma} (\hat{c}^{\dag}_{l,\sigma} \hat{c}_{l+1,\sigma} + {\rm h.c.}) + U \hat{n}_{l,\uparrow} \hat{n}_{l,\downarrow}\right]\ , \quad \label{HamF}
\end{eqnarray}
where $\hat{c}^{\dag}_{l,\sigma}$, $\hat{c}_{l,\sigma}$ and $\hat{n}_{l,\sigma} = \hat{c}^{\dag}_{l,\sigma} \hat{c}_{l,\sigma}$ represent the creation, annihilation and particle number operators of fermion at site-$l$ respectively, $\sigma = \uparrow$, $\downarrow$ denotes the spin polarization, $t > 0$ is the hopping amplitude between two nearest-neighboring sites, and $U$ is the onsite coulomb repulsion.

\begin{figure}[t]
\begin{center}
\includegraphics[width=\linewidth]{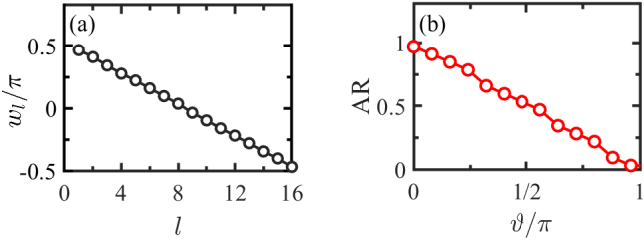}
\caption{(a) In the noninteracting limit $U/t = 0$, we examine the double-even-parity ground state for the Fermi-Hubbard ring (\refeq{HamF}). We train shn-FNN by scheme-$\rm{II}$. In the optimized shn-FNN, we plot the weights $w_l$, which match Eq.~(\refeq{srsF}). (b) When $U/t=0.1$ and $N_{\downarrow}=2$, we plot the accuracy rate $\text{AR}$ of the sign rule $s_{\mathbf{n}_\uparrow,\mathbf{n}_\downarrow} = \text{Sgn}[\cos(\bm{w} \cdot \mathbf{n} + \vartheta)]$ as a function of the variational angle $\vartheta$ for the ground state. The maximum $\text{AR}$ value occurs at $\vartheta=0$. The system size is given by $L=16$.}\label{fig9}
\end{center}
\end{figure}

In the Fock space, each basis is a product of the bases for two species, that is,
\begin{eqnarray}
|\mathbf{n}_\uparrow, \mathbf{n}_\downarrow\rangle = \left[\bigotimes^L_{l=1} | n_{l,\uparrow}\rangle\right] \left[\bigotimes^L_{l=1} | n_{l,\downarrow} \rangle\right]\ .
\end{eqnarray}
Here, we define the vectors $\mathbf{n}_\sigma = (n_{1,\sigma},\ \cdots,\ n_{L,\sigma})$ for species-$\sigma$.
Under the conventional Jordan-Wigner transformation~\cite{Derzhko2001}, the two-channel spin flipping operators $\hat{S}^{\pm}_{l,\sigma}$ can be represented by fermion operators as follows:
\begin{eqnarray}
\begin{split}
\hat{S}^{+}_{l,\sigma} &= \hat{c}^{\dagger}_{l,\sigma} e^{\phantom{-}i \pi \sum_{k<l} \hat{n}_{k,\sigma}}\ ,\\
\hat{S}^{-}_{l,\sigma} &= \hat{c}_{l,\sigma} e^{-i \pi \sum_{k<l} \hat{n}_{k,\sigma}}\ .
\end{split}
\end{eqnarray}
Thus, we get a two-leg spin-$1/2$ ladder
\begin{eqnarray}
\hat{H}_{\text{Ladder}} = \sum_{\sigma} \hat{H}_{\parallel,\sigma} + \hat{H}_{\perp}\ ,
\end{eqnarray}
where
\begin{eqnarray}
& &\hat{H}_{\parallel,\sigma} = \! -t \left[\sum^{L-1}_{l=1} {\hat S}^{+}_{l,\sigma} {\hat S}^{-}_{l+1,\sigma} \! + \! (-1)^{\hat{N}_{\sigma}-1} {\hat S}^{+}_{L,\sigma} {\hat S}^{-}_{1,\sigma} \! + \! {\rm h.c.}\right]\ ,\nonumber\\
& &\hat{H}_{\perp} = U \sum^{L}_{l=1}(\hat{S}^{z}_{l,\uparrow} + 1/2) (\hat{S}^{z}_{l,\downarrow} + 1/2)\ .
\end{eqnarray}
denote the transverse and longitudinal parts, respectively. The particle number operator in total for species-$\sigma$ is given by $\hat{N}_{\sigma} = \sum^L_{l=1} \hat{n}_{l,\sigma}$. We are interested in the ground state for the case of $N_{\uparrow} + N_{\downarrow} = L$, and even $N_{\sigma} = \langle \hat{N}_{\sigma} \rangle$.

\begin{figure}[t]
\begin{center}
\includegraphics[width=\linewidth]{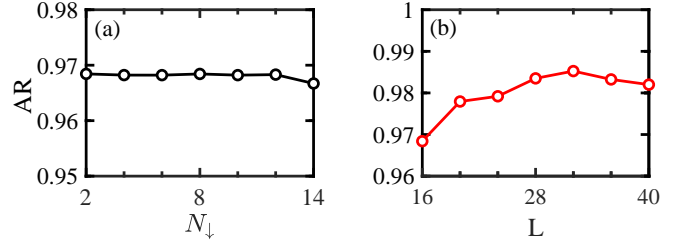}
\caption{(a) The accuracy rate $\text{AR}$ of the sign rule (\refeq{srtF}) as a function of $N_{\downarrow}$ when we choose $L=16$. (b) The same accuracy rate as a function of $L$ when we choose $N_{\downarrow}=2$. Here, we concern the ground state for the Fermi-Hubbard ring (\refeq{HamF}) with $U/t=0.1$.
}\label{fig10}
\end{center}
\end{figure}

When $U=0$, TBC is effectively applied to two decoupled chains in the spin-ladder model, as both $N_{\uparrow}$ and $N_{\downarrow}$ are even. For each species, when the parity of the ground state is even, the optimized shn-FNN with perfect AR can identify the leading-order sign rule given by
\begin{eqnarray}
s^{(e)}_{\mathbf{n}_\sigma} = \text{Sgn}[\cos(\bm{w} \cdot \mathbf{n}_\sigma)]\label{srsF}
\end{eqnarray}
with the weights
\begin{eqnarray}
w_l = -\frac{(l-1)\pi}{L} + \frac{\pi}{2} - \frac{\pi}{2L}\ ,
\end{eqnarray}
depicted in Fig.~\ref{fig9}(a). For the degenerate ground state with odd parity, the function cosine is replaced by the function sine, i.e.,
\begin{eqnarray}
s^{(o)}_{\mathbf{n}_\sigma} = \text{Sgn}[\sin(\bm{w} \cdot \mathbf{n}_\sigma)]\ .
\end{eqnarray}

For small $U/t > 0$ and any $N_{\sigma}$, the parity of the unique ground state is always even. In the case of $U/t=0.1$, $N_{\downarrow} = 2$ and $L=16$, as illustrated in Fig.~\ref{fig9}(b), the optimized shn-FNN suggests a sign rule
\begin{eqnarray}
s_{\mathbf{n}_\uparrow,\mathbf{n}_\downarrow} = \text{Sgn}[\cos(\bm{w} \cdot \mathbf{n})]\ ,\label{srtF}
\end{eqnarray}
with the accuracy rate $\text{AR} \approx 0.97$, where the vector is defined as
\begin{eqnarray}
\mathbf{n} = \mathbf{n}_\uparrow + \mathbf{n}_\downarrow = (n_{1,\uparrow} + n_{1,\downarrow},\ \cdots,\ n_{L,\uparrow} + n_{L,\downarrow})\ .
\end{eqnarray}
So, the resulting leading-order sign rule for the Fermi-Hubbard model remains consistent with the sign rule (\refeq{eq1}).

The leading-order sign rule (\refeq{srtF}) is robust and less dependent on the filling fraction and system size $L$. In the case of $U/t=0.1$ and $L=16$ (Fig.~\ref{fig10}), the accuracy rate $\text{AR}$ is greater than $0.968$ for different $N_\downarrow$. Additionally, as $L$ grows, the accuracy rate $\text{AR}$ for $N_{\downarrow}=2$ gets closer to $0.99$.

\begin{figure}[t]
\begin{center}
\includegraphics[width=\linewidth]{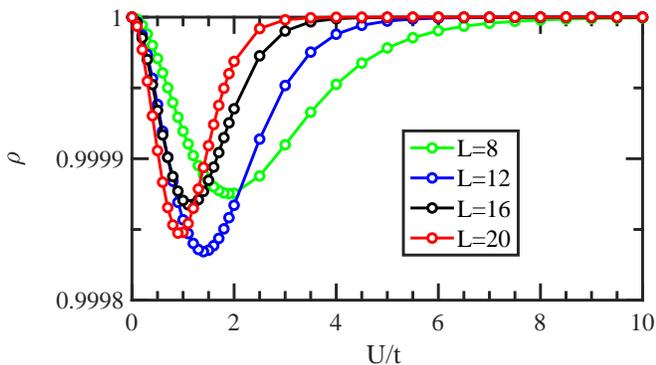}
\caption{The correct weight $\rho$ of the sign rule (\refeq{srtF}) as a function of $U/t$ in the Fermi-Hubbard ring (\refeq{HamF}). We concern the ground state with $N_{\downarrow}=2$, and four different system sizes are investigated: $L=8$ (green), $12$ (blue), $16$ (black), and $20$ (red).
}\label{fig11}
\end{center}
\end{figure}

In the limit of large $U$, only single occupations can exist in the ground state because of a considerable charge gap. As a result, spin fluctuations in the reduced Hilbert space of either spin-up $\hat{c}^{\dagger}_{l,\uparrow} \rvert 0 \rangle$ or spin-down $-\hat{c}^{\dagger}_{l,\downarrow} \rvert 0 \rangle$ are described by the effective antiferromagnetic Heisenberg ring. The ground state for the effective model follows the MPR sign rule exactly. Returning to the fermion bases, it is easy to prove that the weights $w_{l}$ are the same as ones in the leading-order sign rule at $U=0$. We can observe that the corresponding correct weight $\rho$ approaches $1$ when $U/t \ge 8$, as shown in Fig.~\ref{fig11}.

According to the Bethe ansatz solution~\cite{Lieb1968}, the Fermi liquid only survives at $U=0$ in the thermodynamical limit (TDL). However, because of a tiny charge gap close to $U=0$, fermions exhibit behavior like a Fermi liquid in the ground state for system size limited to $L \le 20$, much smaller than the correlation length. Consequently, a quasi-critical point is indicated by the minimum of the correct weight $\rho$, where the strong quantum fluctuations would strongly violate the leading-order sign rules. As $L$ grows in Fig.~\ref{fig11}, the quasi-critical point gradually approaches $U=0$.

In an alternative definition of bases, that is,
\begin{eqnarray}
\rvert \mathbf{n} \rangle = \bigotimes^{L}_{l=1} \bigg{[} \rvert n_{l,\uparrow} \rangle \rvert n_{l,\downarrow} \rangle \bigg{]}\ ,
\end{eqnarray}
the Jordan Wigner transformation changes accordingly, and an additional nonlinear appendix
\begin{eqnarray}
(-1)^{\sum^L_{l=2} n_{l,\uparrow} \sum^{l-1}_{k=1} n_{k,\downarrow}}
\end{eqnarray}
exists in front of the predicted sign rules. However, this appendix can not be expressed in shn-FNN.

\section{Summary and discussions}\label{secE}
We have successfully developed a Gutzwiller mean-field theory of sign rules for the ordered ground states in qubit lattice models, which perfectly matches the sign predicted by a shallow FNN with a single hidden neuron, called shn-FNN. By utilizing this principle, we provide a consistent explanation for the excellent performance of activation functions in the neural network and offer a vivid interpretation of the sign rule represented by FNN.

We systematically test our theory on various spin models and the Fermi-Hubbard ring. For non-frustrated spin-$1/2$ models, such as a generalized Ising ring, (twisted) XY rings, and an antiferromagnetic Heisenberg ring, the sign rules for ground states with magnetic orders can be fully captured by shn-FNN, where the accurate rate of the prediction can archive $1$ exactly. However, in the case of frustrated models where interactions compete, the complexity of sign rules for ground states is significantly enhanced, reducing prediction accuracy. Nonetheless, the leading-order sign rules obtained by optimizing shn-FNN still provide a visual scenario of orders in spins, with the characteristic weight vector closely related to pitch angles. In the Fermi-Hubbard ring, we can obtain a unified sign rule by selecting suitable bases.

GWMF may not be suitable for 1D models since quantum fluctuations tend to destroy long-range orders. However, our current work presents a fresh perspective by demonstrating that GWMF can effectively capture the leading-order sign rule in the wave function, where fluctuations in amplitudes are erased. Our theory is a simple starting point by removing short-range details in ordered states. It would be intriguing to explore the information encoded in high-order microscopic processes instead of focusing solely on the leading-order ones. Of course, the theory for general lattice models also deserves profound studies in the future.

\section{Acknowlegement}\label{secF}
We thank Tao Li, Rui Wang, Ji-Lu He, Wei Su, and Wei Pan for the grateful discussion. S. H. acknowledges funding from the Ministry of Science and Technology of China (Grant No.~2022YFA1402700) and the National Science Foundation of China (Grants No.~12174020). Z. P. Y. acknowledges funding from the National Science
Foundation of China (Grants No.~12074041). S. H. and K. X. further acknowledge support from Grant NSAF-U2230402. The computations were performed on the Tianhe-2JK at the Beijing Computational Science Research Center (CSRC) and the high-performance computing cluster of Beijing Normal University in Zhuhai.

\vspace{0.5cm}

\appendix

\section{Sign rule for the twisted ferromagnetic spin-$1/2$ XY rings}\label{appC}
Here we prove that the even-parity ground-state wave function for the Hamiltonian $\hat{H}^P_\text{xy} (J\tilde{J})$ with $J < 0$, mentioned in Sec.~\ref{subsecA3} of the main text, has positive signs $s_\mathbf{n} > 0$.

Under the Jordan-Wigner transformation~\cite{Derzhko2001}, the spin operators can be represented by fermion operators as follows:
\begin{eqnarray}
\hat{S}^{+}_l = \hat{c}^{\dagger}_l e^{i \pi \sum_{k<l} \hat{n}_k} \ \text{and} \ \hat{S}^{-}_l = \hat{c}_l e^{-i \pi \sum_{k<l} \hat{n}_k}\ 
\end{eqnarray}
where $\hat{c}_l$, $\hat{c}^{\dagger}_l$, and $\hat{n}_l = \hat{c}^\dag_l \hat{c}_l$ represent the annihilation, creation, and particle number operators of fermion, respectively. As a result, the Hamiltonian $\hat{H}^P_\text{xy} (J\tilde{J})$ can be transformed into the one for spinless free fermions defined as
\begin{eqnarray}
\hat{H}^{P/T}_F (J) = J \left(\sum^{L-1}_{l=1} \hat{c}^\dag_l \hat{c}_{l+1}\pm \hat{c}^\dag_L \hat{c}_1 \right) + {\rm h.c.}.
\end{eqnarray}
The selection of periodic or twisted boundary conditions in the fermion model depends on whether the particle number $N$ is odd or even.

For odd $N$, the single-particle levels are described by plane waves $\exp(i k_m l)$ with discrete momenta $k_m = 2m \pi / L$, where the integer $m$ ranges from $0$ to $L-1$. The energy $\epsilon_m$ for the $m$-th single-particle level follows a formula $\epsilon_m = 2 J \cos(k_m + \varphi)$ with the phase angle $\varphi = \pi / L$. At half-filling $N = L / 2$, the ground state selects single-particle levels with the integer $m \in (-L / 4,\ L / 4]$. Therefore, the ground-state wave function for the Hamiltonian $\hat{H}^P_F (J\tilde{J})$, seen as a determinant of selected plane waves, is the same as the one for the other Hamiltonian $\hat{H}^P_F (J)$ at half-filling. Similarly, when $N$ is even, the ground-state wave function for the Hamiltonian $\hat{H}^P_F (J\tilde{J})$ is the same as the one for the Hamiltonian $\hat{H}^T_F (J)$ as well. In conclusion, both Hamiltonian $\hat{H}^{P/T}_F (J)$ for odd/even $N$ can be transformed back to the unique Hamiltonian $\hat{H}^P_\text{xy} (J)$ through the inverse Jordan-Wigner transformation, where the even-parity ground-state wave function always has positive signs, according to the Perron-Frobenius theorem~\cite{Perron1907, Frobenius1909}.

\bibliography{refprb}

%merlin.mbs apsrev4-1.bst 2010-07-25 4.21a (PWD, AO, DPC) hacked
%Control: key (0)
%Control: author (72) initials jnrlst
%Control: editor formatted (1) identically to author
%Control: production of article title (-1) disabled
%Control: page (0) single
%Control: year (1) truncated
%Control: production of eprint (0) enabled
\begin{thebibliography}{68}%
\makeatletter
\providecommand \@ifxundefined [1]{%
 \@ifx{#1\undefined}
}%
\providecommand \@ifnum [1]{%
 \ifnum #1\expandafter \@firstoftwo
 \else \expandafter \@secondoftwo
 \fi
}%
\providecommand \@ifx [1]{%
 \ifx #1\expandafter \@firstoftwo
 \else \expandafter \@secondoftwo
 \fi
}%
\providecommand \natexlab [1]{#1}%
\providecommand \enquote  [1]{``#1''}%
\providecommand \bibnamefont  [1]{#1}%
\providecommand \bibfnamefont [1]{#1}%
\providecommand \citenamefont [1]{#1}%
\providecommand \href@noop [0]{\@secondoftwo}%
\providecommand \href [0]{\begingroup \@sanitize@url \@href}%
\providecommand \@href[1]{\@@startlink{#1}\@@href}%
\providecommand \@@href[1]{\endgroup#1\@@endlink}%
\providecommand \@sanitize@url [0]{\catcode `\\12\catcode `\$12\catcode
  `\&12\catcode `\#12\catcode `\^12\catcode `\_12\catcode `\%12\relax}%
\providecommand \@@startlink[1]{}%
\providecommand \@@endlink[0]{}%
\providecommand \url  [0]{\begingroup\@sanitize@url \@url }%
\providecommand \@url [1]{\endgroup\@href {#1}{\urlprefix }}%
\providecommand \urlprefix  [0]{URL }%
\providecommand \Eprint [0]{\href }%
\providecommand \doibase [0]{http://dx.doi.org/}%
\providecommand \selectlanguage [0]{\@gobble}%
\providecommand \bibinfo  [0]{\@secondoftwo}%
\providecommand \bibfield  [0]{\@secondoftwo}%
\providecommand \translation [1]{[#1]}%
\providecommand \BibitemOpen [0]{}%
\providecommand \bibitemStop [0]{}%
\providecommand \bibitemNoStop [0]{.\EOS\space}%
\providecommand \EOS [0]{\spacefactor3000\relax}%
\providecommand \BibitemShut  [1]{\csname bibitem#1\endcsname}%
\let\auto@bib@innerbib\@empty
%</preamble>
\bibitem [{\citenamefont {Eisert}\ \emph {et~al.}(2010)\citenamefont {Eisert},
  \citenamefont {Cramer},\ and\ \citenamefont {Plenio}}]{Eisert2010}%
  \BibitemOpen
  \bibfield  {author} {\bibinfo {author} {\bibfnamefont {J.}~\bibnamefont
  {Eisert}}, \bibinfo {author} {\bibfnamefont {M.}~\bibnamefont {Cramer}}, \
  and\ \bibinfo {author} {\bibfnamefont {M.~B.}\ \bibnamefont {Plenio}},\
  }\href {\doibase 10.1103/revmodphys.82.277} {\bibfield  {journal} {\bibinfo
  {journal} {Reviews of Modern Physics}\ }\textbf {\bibinfo {volume} {82}},\
  \bibinfo {pages} {277} (\bibinfo {year} {2010})}\BibitemShut {NoStop}%
\bibitem [{\citenamefont {Chertkov}\ and\ \citenamefont
  {Clark}(2018)}]{Chertkov2018}%
  \BibitemOpen
  \bibfield  {author} {\bibinfo {author} {\bibfnamefont {E.}~\bibnamefont
  {Chertkov}}\ and\ \bibinfo {author} {\bibfnamefont {B.~K.}\ \bibnamefont
  {Clark}},\ }\href {\doibase 10.1103/physrevx.8.031029} {\bibfield  {journal}
  {\bibinfo  {journal} {Physical Review X}\ }\textbf {\bibinfo {volume} {8}},\
  \bibinfo {pages} {031029} (\bibinfo {year} {2018})}\BibitemShut {NoStop}%
\bibitem [{\citenamefont {Wang}\ \emph {et~al.}(2019)\citenamefont {Wang},
  \citenamefont {Zhai},\ and\ \citenamefont {You}}]{Wang2019}%
  \BibitemOpen
  \bibfield  {author} {\bibinfo {author} {\bibfnamefont {C.}~\bibnamefont
  {Wang}}, \bibinfo {author} {\bibfnamefont {H.}~\bibnamefont {Zhai}}, \ and\
  \bibinfo {author} {\bibfnamefont {Y.~Z.}\ \bibnamefont {You}},\ }\href
  {\doibase 10.1016/j.scib.2019.07.014} {\bibfield  {journal} {\bibinfo
  {journal} {Science Bulletin}\ }\textbf {\bibinfo {volume} {64}},\ \bibinfo
  {pages} {1228} (\bibinfo {year} {2019})}\BibitemShut {NoStop}%
\bibitem [{\citenamefont {Irkhin}\ and\ \citenamefont
  {Skryabin}(2019)}]{Irkhin2019}%
  \BibitemOpen
  \bibfield  {author} {\bibinfo {author} {\bibfnamefont {V.~Y.}\ \bibnamefont
  {Irkhin}}\ and\ \bibinfo {author} {\bibfnamefont {Y.~N.}\ \bibnamefont
  {Skryabin}},\ }\href {\doibase 10.1134/s0031918x19060061} {\bibfield
  {journal} {\bibinfo  {journal} {Physics of Metals and Metallography}\
  }\textbf {\bibinfo {volume} {120}},\ \bibinfo {pages} {513} (\bibinfo {year}
  {2019})}\BibitemShut {NoStop}%
\bibitem [{\citenamefont {Grover}\ and\ \citenamefont
  {Fisher}(2015)}]{Grover2015}%
  \BibitemOpen
  \bibfield  {author} {\bibinfo {author} {\bibfnamefont {T.}~\bibnamefont
  {Grover}}\ and\ \bibinfo {author} {\bibfnamefont {M.~P.~A.}\ \bibnamefont
  {Fisher}},\ }\href {\doibase 10.1103/physreva.92.042308} {\bibfield
  {journal} {\bibinfo  {journal} {Physical Review A}\ }\textbf {\bibinfo
  {volume} {92}},\ \bibinfo {pages} {042308} (\bibinfo {year}
  {2015})}\BibitemShut {NoStop}%
\bibitem [{\citenamefont {Perron}(1907)}]{Perron1907}%
  \BibitemOpen
  \bibfield  {author} {\bibinfo {author} {\bibfnamefont {O.}~\bibnamefont
  {Perron}},\ }\href {\doibase 10.1007/bf01449896} {\bibfield  {journal}
  {\bibinfo  {journal} {Mathematische Annalen}\ }\textbf {\bibinfo {volume}
  {64}},\ \bibinfo {pages} {248} (\bibinfo {year} {1907})}\BibitemShut
  {NoStop}%
\bibitem [{\citenamefont {Frobenius}(1909)}]{Frobenius1909}%
  \BibitemOpen
  \bibfield  {author} {\bibinfo {author} {\bibfnamefont {G.}~\bibnamefont
  {Frobenius}},\ }\href@noop {} {\emph {\bibinfo {title} {Matrices from
  positive elements. II}}},\ \bibinfo {number} {1}\ (\bibinfo  {publisher}
  {Walter de Gruyter \& CO},\ \bibinfo {address} {Berlin, Germany},\ \bibinfo
  {year} {1909})\ pp.\ \bibinfo {pages} {514--518}\BibitemShut {NoStop}%
\bibitem [{\citenamefont {Marshall}(1955)}]{Marshall1955}%
  \BibitemOpen
  \bibfield  {author} {\bibinfo {author} {\bibfnamefont {W.}~\bibnamefont
  {Marshall}},\ }\href {\doibase 10.1098/rspa.1955.0200} {\bibfield  {journal}
  {\bibinfo  {journal} {Proceedings of the Royal Society of London. Series A.
  Mathematical and Physical Sciences}\ }\textbf {\bibinfo {volume} {232}},\
  \bibinfo {pages} {48} (\bibinfo {year} {1955})}\BibitemShut {NoStop}%
\bibitem [{\citenamefont {Lieb}\ \emph {et~al.}(1961)\citenamefont {Lieb},
  \citenamefont {Schultz},\ and\ \citenamefont {Mattis}}]{Lieb1961}%
  \BibitemOpen
  \bibfield  {author} {\bibinfo {author} {\bibfnamefont {E.}~\bibnamefont
  {Lieb}}, \bibinfo {author} {\bibfnamefont {T.}~\bibnamefont {Schultz}}, \
  and\ \bibinfo {author} {\bibfnamefont {D.}~\bibnamefont {Mattis}},\ }\href
  {\doibase 10.1016/0003-4916(61)90115-4} {\bibfield  {journal} {\bibinfo
  {journal} {Annals of Physics}\ }\textbf {\bibinfo {volume} {16}},\ \bibinfo
  {pages} {407} (\bibinfo {year} {1961})}\BibitemShut {NoStop}%
\bibitem [{\citenamefont {Lieb}\ and\ \citenamefont {Mattis}(1962)}]{Lieb1962}%
  \BibitemOpen
  \bibfield  {author} {\bibinfo {author} {\bibfnamefont {E.}~\bibnamefont
  {Lieb}}\ and\ \bibinfo {author} {\bibfnamefont {D.}~\bibnamefont {Mattis}},\
  }\href {\doibase 10.1063/1.1724276} {\bibfield  {journal} {\bibinfo
  {journal} {Journal of Mathematical Physics}\ }\textbf {\bibinfo {volume}
  {3}},\ \bibinfo {pages} {749} (\bibinfo {year} {1962})}\BibitemShut {NoStop}%
\bibitem [{\citenamefont {Zeng}\ and\ \citenamefont
  {Parkinson}(1995)}]{Zeng1995}%
  \BibitemOpen
  \bibfield  {author} {\bibinfo {author} {\bibfnamefont {C.}~\bibnamefont
  {Zeng}}\ and\ \bibinfo {author} {\bibfnamefont {J.~B.}\ \bibnamefont
  {Parkinson}},\ }\href {\doibase 10.1103/PhysRevB.51.11609} {\bibfield
  {journal} {\bibinfo  {journal} {Phys. Rev. B}\ }\textbf {\bibinfo {volume}
  {51}},\ \bibinfo {pages} {11609} (\bibinfo {year} {1995})}\BibitemShut
  {NoStop}%
\bibitem [{\citenamefont {Bursill}\ \emph {et~al.}(1995)\citenamefont
  {Bursill}, \citenamefont {Gehring}, \citenamefont {Farnell}, \citenamefont
  {Parkinson}, \citenamefont {Xiang},\ and\ \citenamefont
  {Zeng}}]{Bursill1995}%
  \BibitemOpen
  \bibfield  {author} {\bibinfo {author} {\bibfnamefont {R.}~\bibnamefont
  {Bursill}}, \bibinfo {author} {\bibfnamefont {G.~A.}\ \bibnamefont
  {Gehring}}, \bibinfo {author} {\bibfnamefont {D.~J.~J.}\ \bibnamefont
  {Farnell}}, \bibinfo {author} {\bibfnamefont {J.~B.}\ \bibnamefont
  {Parkinson}}, \bibinfo {author} {\bibfnamefont {T.}~\bibnamefont {Xiang}}, \
  and\ \bibinfo {author} {\bibfnamefont {C.}~\bibnamefont {Zeng}},\ }\href
  {\doibase 10.1088/0953-8984/7/45/016} {\bibfield  {journal} {\bibinfo
  {journal} {Journal of Physics: Condensed Matter}\ }\textbf {\bibinfo {volume}
  {7}},\ \bibinfo {pages} {8605} (\bibinfo {year} {1995})}\BibitemShut
  {NoStop}%
\bibitem [{\citenamefont {Retzlaff}\ \emph {et~al.}(1993)\citenamefont
  {Retzlaff}, \citenamefont {Richter},\ and\ \citenamefont
  {Ivanov}}]{Retzla1993}%
  \BibitemOpen
  \bibfield  {author} {\bibinfo {author} {\bibfnamefont {K.}~\bibnamefont
  {Retzlaff}}, \bibinfo {author} {\bibfnamefont {J.}~\bibnamefont {Richter}}, \
  and\ \bibinfo {author} {\bibfnamefont {N.~B.}\ \bibnamefont {Ivanov}},\
  }\href {\doibase 10.1007/bf01308803} {\bibfield  {journal} {\bibinfo
  {journal} {Zeitschrift f{\"u}r Physik B Condensed Matter}\ }\textbf {\bibinfo
  {volume} {93}},\ \bibinfo {pages} {21} (\bibinfo {year} {1993})}\BibitemShut
  {NoStop}%
\bibitem [{\citenamefont {Richter}\ \emph {et~al.}(1994)\citenamefont
  {Richter}, \citenamefont {Ivanov},\ and\ \citenamefont
  {Retzlaff}}]{Richte1994}%
  \BibitemOpen
  \bibfield  {author} {\bibinfo {author} {\bibfnamefont {J.}~\bibnamefont
  {Richter}}, \bibinfo {author} {\bibfnamefont {N.~B.}\ \bibnamefont {Ivanov}},
  \ and\ \bibinfo {author} {\bibfnamefont {K.}~\bibnamefont {Retzlaff}},\
  }\href {\doibase 10.1209/0295-5075/25/7/012} {\bibfield  {journal} {\bibinfo
  {journal} {Europhysics Letters}\ }\textbf {\bibinfo {volume} {25}},\ \bibinfo
  {pages} {545} (\bibinfo {year} {1994})}\BibitemShut {NoStop}%
\bibitem [{\citenamefont {Cai}\ and\ \citenamefont {Liu}(2018)}]{Cai2018}%
  \BibitemOpen
  \bibfield  {author} {\bibinfo {author} {\bibfnamefont {Z.}~\bibnamefont
  {Cai}}\ and\ \bibinfo {author} {\bibfnamefont {J.}~\bibnamefont {Liu}},\
  }\href {\doibase 10.1103/PhysRevB.97.035116} {\bibfield  {journal} {\bibinfo
  {journal} {Phys. Rev. B}\ }\textbf {\bibinfo {volume} {97}},\ \bibinfo
  {pages} {035116} (\bibinfo {year} {2018})}\BibitemShut {NoStop}%
\bibitem [{\citenamefont {Westerhout}\ \emph {et~al.}(2020)\citenamefont
  {Westerhout}, \citenamefont {Astrakhantsev}, \citenamefont {Tikhonov},
  \citenamefont {Katsnelson},\ and\ \citenamefont {Bagrov}}]{Westerhout2020}%
  \BibitemOpen
  \bibfield  {author} {\bibinfo {author} {\bibfnamefont {T.}~\bibnamefont
  {Westerhout}}, \bibinfo {author} {\bibfnamefont {N.}~\bibnamefont
  {Astrakhantsev}}, \bibinfo {author} {\bibfnamefont {K.~S.}\ \bibnamefont
  {Tikhonov}}, \bibinfo {author} {\bibfnamefont {M.~I.}\ \bibnamefont
  {Katsnelson}}, \ and\ \bibinfo {author} {\bibfnamefont {A.~A.}\ \bibnamefont
  {Bagrov}},\ }\href {\doibase 10.1038/s41467-020-15402-w} {\bibfield
  {journal} {\bibinfo  {journal} {Nature Communications}\ }\textbf {\bibinfo
  {volume} {11}},\ \bibinfo {pages} {1593} (\bibinfo {year}
  {2020})}\BibitemShut {NoStop}%
\bibitem [{\citenamefont {White}(1992)}]{White1992}%
  \BibitemOpen
  \bibfield  {author} {\bibinfo {author} {\bibfnamefont {S.~R.}\ \bibnamefont
  {White}},\ }\href {\doibase 10.1103/physrevlett.69.2863} {\bibfield
  {journal} {\bibinfo  {journal} {Physical Review Letters}\ }\textbf {\bibinfo
  {volume} {69}},\ \bibinfo {pages} {2863} (\bibinfo {year}
  {1992})}\BibitemShut {NoStop}%
\bibitem [{\citenamefont {Peschel}\ \emph {et~al.}(1999)\citenamefont
  {Peschel}, \citenamefont {Kaulke}, \citenamefont {Wang},\ and\ \citenamefont
  {Hallberg}}]{Wang1999}%
  \BibitemOpen
  \bibinfo {editor} {\bibfnamefont {I.}~\bibnamefont {Peschel}}, \bibinfo
  {editor} {\bibfnamefont {M.}~\bibnamefont {Kaulke}}, \bibinfo {editor}
  {\bibfnamefont {X.}~\bibnamefont {Wang}}, \ and\ \bibinfo {editor}
  {\bibfnamefont {K.}~\bibnamefont {Hallberg}},\ eds.,\ \href {\doibase
  10.1007/bfb0106062} {\emph {\bibinfo {title} {Density-Matrix
  Renormalization}}}\ (\bibinfo  {publisher} {Springer Berlin Heidelberg},\
  \bibinfo {year} {1999})\BibitemShut {NoStop}%
\bibitem [{\citenamefont {Schollw{\"o}ck}(2011)}]{Schollwock2011}%
  \BibitemOpen
  \bibfield  {author} {\bibinfo {author} {\bibfnamefont {U.}~\bibnamefont
  {Schollw{\"o}ck}},\ }\href {\doibase 10.1016/j.aop.2010.09.012} {\bibfield
  {journal} {\bibinfo  {journal} {Annals of Physics}\ }\textbf {\bibinfo
  {volume} {326}},\ \bibinfo {pages} {96} (\bibinfo {year} {2011})}\BibitemShut
  {NoStop}%
\bibitem [{\citenamefont {Or{\'{u}}s}(2014)}]{Orus2014}%
  \BibitemOpen
  \bibfield  {author} {\bibinfo {author} {\bibfnamefont {R.}~\bibnamefont
  {Or{\'{u}}s}},\ }\href {\doibase 10.1016/j.aop.2014.06.013} {\bibfield
  {journal} {\bibinfo  {journal} {Annals of Physics}\ }\textbf {\bibinfo
  {volume} {349}},\ \bibinfo {pages} {117} (\bibinfo {year}
  {2014})}\BibitemShut {NoStop}%
\bibitem [{\citenamefont {Carleo}\ and\ \citenamefont
  {Troyer}(2017)}]{Carleo2017}%
  \BibitemOpen
  \bibfield  {author} {\bibinfo {author} {\bibfnamefont {G.}~\bibnamefont
  {Carleo}}\ and\ \bibinfo {author} {\bibfnamefont {M.}~\bibnamefont
  {Troyer}},\ }\href {\doibase 10.1126/science.aag2302} {\bibfield  {journal}
  {\bibinfo  {journal} {Science}\ }\textbf {\bibinfo {volume} {355}},\ \bibinfo
  {pages} {602} (\bibinfo {year} {2017})}\BibitemShut {NoStop}%
\bibitem [{\citenamefont {Carleo}\ \emph {et~al.}(2019)\citenamefont {Carleo},
  \citenamefont {Cirac}, \citenamefont {Cranmer}, \citenamefont {Daudet},
  \citenamefont {Schuld}, \citenamefont {Tishby}, \citenamefont
  {Vogt-Maranto},\ and\ \citenamefont {Zdeborov{\'{a}}}}]{Carleo2019}%
  \BibitemOpen
  \bibfield  {author} {\bibinfo {author} {\bibfnamefont {G.}~\bibnamefont
  {Carleo}}, \bibinfo {author} {\bibfnamefont {I.}~\bibnamefont {Cirac}},
  \bibinfo {author} {\bibfnamefont {K.}~\bibnamefont {Cranmer}}, \bibinfo
  {author} {\bibfnamefont {L.}~\bibnamefont {Daudet}}, \bibinfo {author}
  {\bibfnamefont {M.}~\bibnamefont {Schuld}}, \bibinfo {author} {\bibfnamefont
  {N.}~\bibnamefont {Tishby}}, \bibinfo {author} {\bibfnamefont
  {L.}~\bibnamefont {Vogt-Maranto}}, \ and\ \bibinfo {author} {\bibfnamefont
  {L.}~\bibnamefont {Zdeborov{\'{a}}}},\ }\href {\doibase
  10.1103/revmodphys.91.045002} {\bibfield  {journal} {\bibinfo  {journal}
  {Reviews of Modern Physics}\ }\textbf {\bibinfo {volume} {91}},\ \bibinfo
  {pages} {045002} (\bibinfo {year} {2019})}\BibitemShut {NoStop}%
\bibitem [{\citenamefont {Jia}\ \emph {et~al.}(2019)\citenamefont {Jia},
  \citenamefont {Yi}, \citenamefont {Zhai}, \citenamefont {Wu}, \citenamefont
  {Guo},\ and\ \citenamefont {Guo}}]{Jia2019}%
  \BibitemOpen
  \bibfield  {author} {\bibinfo {author} {\bibfnamefont {Z.~A.}\ \bibnamefont
  {Jia}}, \bibinfo {author} {\bibfnamefont {B.}~\bibnamefont {Yi}}, \bibinfo
  {author} {\bibfnamefont {R.}~\bibnamefont {Zhai}}, \bibinfo {author}
  {\bibfnamefont {Y.~C.}\ \bibnamefont {Wu}}, \bibinfo {author} {\bibfnamefont
  {G.~C.}\ \bibnamefont {Guo}}, \ and\ \bibinfo {author} {\bibfnamefont
  {G.~P.}\ \bibnamefont {Guo}},\ }\href {\doibase 10.1002/qute.201800077}
  {\bibfield  {journal} {\bibinfo  {journal} {Advanced Quantum Technologies}\
  }\textbf {\bibinfo {volume} {2}},\ \bibinfo {pages} {1800077} (\bibinfo
  {year} {2019})}\BibitemShut {NoStop}%
\bibitem [{\citenamefont {Vivas}\ \emph {et~al.}(2022)\citenamefont {Vivas},
  \citenamefont {Madro{\~n}ero}, \citenamefont {Bucheli}, \citenamefont
  {G{\'o}mez},\ and\ \citenamefont {Reina}}]{Vivas2022}%
  \BibitemOpen
  \bibfield  {author} {\bibinfo {author} {\bibfnamefont {D.~R.}\ \bibnamefont
  {Vivas}}, \bibinfo {author} {\bibfnamefont {J.}~\bibnamefont
  {Madro{\~n}ero}}, \bibinfo {author} {\bibfnamefont {V.}~\bibnamefont
  {Bucheli}}, \bibinfo {author} {\bibfnamefont {L.~O.}\ \bibnamefont
  {G{\'o}mez}}, \ and\ \bibinfo {author} {\bibfnamefont {J.~H.}\ \bibnamefont
  {Reina}},\ }\href@noop {} {\bibfield  {journal} {\bibinfo  {journal} {arXiv
  preprint arXiv:2204.12966}\ } (\bibinfo {year} {2022})}\BibitemShut {NoStop}%
\bibitem [{\citenamefont {Choo}\ \emph {et~al.}(2019)\citenamefont {Choo},
  \citenamefont {Neupert},\ and\ \citenamefont {Carleo}}]{Choo2019}%
  \BibitemOpen
  \bibfield  {author} {\bibinfo {author} {\bibfnamefont {K.}~\bibnamefont
  {Choo}}, \bibinfo {author} {\bibfnamefont {T.}~\bibnamefont {Neupert}}, \
  and\ \bibinfo {author} {\bibfnamefont {G.}~\bibnamefont {Carleo}},\ }\href
  {\doibase 10.1103/physrevb.100.125124} {\bibfield  {journal} {\bibinfo
  {journal} {Physical Review B}\ }\textbf {\bibinfo {volume} {100}},\ \bibinfo
  {pages} {125124} (\bibinfo {year} {2019})}\BibitemShut {NoStop}%
\bibitem [{\citenamefont {Szab{\'{o}}}\ and\ \citenamefont
  {Castelnovo}(2020)}]{Szabo2020}%
  \BibitemOpen
  \bibfield  {author} {\bibinfo {author} {\bibfnamefont {A.}~\bibnamefont
  {Szab{\'{o}}}}\ and\ \bibinfo {author} {\bibfnamefont {C.}~\bibnamefont
  {Castelnovo}},\ }\href {\doibase 10.1103/physrevresearch.2.033075} {\bibfield
   {journal} {\bibinfo  {journal} {Physical Review Research}\ }\textbf
  {\bibinfo {volume} {2}},\ \bibinfo {pages} {033075} (\bibinfo {year}
  {2020})}\BibitemShut {NoStop}%
\bibitem [{\citenamefont {Bukov}\ \emph {et~al.}(2021)\citenamefont {Bukov},
  \citenamefont {Schmitt},\ and\ \citenamefont {Dupont}}]{Bukov2021}%
  \BibitemOpen
  \bibfield  {author} {\bibinfo {author} {\bibfnamefont {M.}~\bibnamefont
  {Bukov}}, \bibinfo {author} {\bibfnamefont {M.}~\bibnamefont {Schmitt}}, \
  and\ \bibinfo {author} {\bibfnamefont {M.}~\bibnamefont {Dupont}},\ }\href
  {\doibase 10.21468/scipostphys.10.6.147} {\bibfield  {journal} {\bibinfo
  {journal} {{SciPost} Physics}\ }\textbf {\bibinfo {volume} {10}},\ \bibinfo
  {pages} {147} (\bibinfo {year} {2021})}\BibitemShut {NoStop}%
\bibitem [{\citenamefont {Roscher}\ \emph {et~al.}(2020)\citenamefont
  {Roscher}, \citenamefont {Bohn}, \citenamefont {Duarte},\ and\ \citenamefont
  {Garcke}}]{Roscher2020}%
  \BibitemOpen
  \bibfield  {author} {\bibinfo {author} {\bibfnamefont {R.}~\bibnamefont
  {Roscher}}, \bibinfo {author} {\bibfnamefont {B.}~\bibnamefont {Bohn}},
  \bibinfo {author} {\bibfnamefont {M.~F.}\ \bibnamefont {Duarte}}, \ and\
  \bibinfo {author} {\bibfnamefont {J.}~\bibnamefont {Garcke}},\ }\href
  {\doibase 10.1109/ACCESS.2020.2976199} {\bibfield  {journal} {\bibinfo
  {journal} {IEEE Access}\ }\textbf {\bibinfo {volume} {8}},\ \bibinfo {pages}
  {42200} (\bibinfo {year} {2020})}\BibitemShut {NoStop}%
\bibitem [{\citenamefont {He}\ \emph {et~al.}(2020)\citenamefont {He},
  \citenamefont {Ma},\ and\ \citenamefont {Wang}}]{He2020}%
  \BibitemOpen
  \bibfield  {author} {\bibinfo {author} {\bibfnamefont {C.}~\bibnamefont
  {He}}, \bibinfo {author} {\bibfnamefont {M.}~\bibnamefont {Ma}}, \ and\
  \bibinfo {author} {\bibfnamefont {P.}~\bibnamefont {Wang}},\ }\href {\doibase
  10.1016/j.neucom.2020.01.036} {\bibfield  {journal} {\bibinfo  {journal}
  {Neurocomputing}\ }\textbf {\bibinfo {volume} {387}},\ \bibinfo {pages} {346}
  (\bibinfo {year} {2020})}\BibitemShut {NoStop}%
\bibitem [{\citenamefont {Fan}\ \emph {et~al.}(2021)\citenamefont {Fan},
  \citenamefont {Xiong}, \citenamefont {Li},\ and\ \citenamefont
  {Wang}}]{Fan2021}%
  \BibitemOpen
  \bibfield  {author} {\bibinfo {author} {\bibfnamefont {F.~L.}\ \bibnamefont
  {Fan}}, \bibinfo {author} {\bibfnamefont {J.~J.}\ \bibnamefont {Xiong}},
  \bibinfo {author} {\bibfnamefont {M.~Z.}\ \bibnamefont {Li}}, \ and\ \bibinfo
  {author} {\bibfnamefont {G.}~\bibnamefont {Wang}},\ }\href {\doibase
  10.1109/TRPMS.2021.3066428} {\bibfield  {journal} {\bibinfo  {journal} {IEEE
  Transactions on Radiation and Plasma Medical Sciences}\ }\textbf {\bibinfo
  {volume} {5}},\ \bibinfo {pages} {741} (\bibinfo {year} {2021})}\BibitemShut
  {NoStop}%
\bibitem [{\citenamefont {Raissi}\ \emph {et~al.}(2019)\citenamefont {Raissi},
  \citenamefont {Perdikaris},\ and\ \citenamefont {Karniadakis}}]{Raissi2019}%
  \BibitemOpen
  \bibfield  {author} {\bibinfo {author} {\bibfnamefont {M.}~\bibnamefont
  {Raissi}}, \bibinfo {author} {\bibfnamefont {P.}~\bibnamefont {Perdikaris}},
  \ and\ \bibinfo {author} {\bibfnamefont {G.}~\bibnamefont {Karniadakis}},\
  }\href {\doibase 10.1016/j.jcp.2018.10.045} {\bibfield  {journal} {\bibinfo
  {journal} {Journal of Computational Physics}\ }\textbf {\bibinfo {volume}
  {378}},\ \bibinfo {pages} {686} (\bibinfo {year} {2019})}\BibitemShut
  {NoStop}%
\bibitem [{\citenamefont {Yuan}\ and\ \citenamefont {Weng}(2021)}]{Yuan2021}%
  \BibitemOpen
  \bibfield  {author} {\bibinfo {author} {\bibfnamefont {J.}~\bibnamefont
  {Yuan}}\ and\ \bibinfo {author} {\bibfnamefont {Y.}~\bibnamefont {Weng}},\
  }in\ \href {\doibase 10.1109/icdm51629.2021.00096} {\emph {\bibinfo
  {booktitle} {2021 {IEEE} International Conference on Data Mining ({ICDM})}}}\
  (\bibinfo  {publisher} {{IEEE}},\ \bibinfo {year} {2021})\BibitemShut
  {NoStop}%
\bibitem [{\citenamefont {Cai}\ \emph {et~al.}(2021)\citenamefont {Cai},
  \citenamefont {Mao}, \citenamefont {Wang}, \citenamefont {Yin},\ and\
  \citenamefont {Karniadakis}}]{Cai2021}%
  \BibitemOpen
  \bibfield  {author} {\bibinfo {author} {\bibfnamefont {S.}~\bibnamefont
  {Cai}}, \bibinfo {author} {\bibfnamefont {Z.}~\bibnamefont {Mao}}, \bibinfo
  {author} {\bibfnamefont {Z.}~\bibnamefont {Wang}}, \bibinfo {author}
  {\bibfnamefont {M.}~\bibnamefont {Yin}}, \ and\ \bibinfo {author}
  {\bibfnamefont {G.~E.}\ \bibnamefont {Karniadakis}},\ }\href {\doibase
  10.1007/s10409-021-01148-1} {\bibfield  {journal} {\bibinfo  {journal} {Acta
  Mechanica Sinica}\ }\textbf {\bibinfo {volume} {37}},\ \bibinfo {pages}
  {1727} (\bibinfo {year} {2021})}\BibitemShut {NoStop}%
\bibitem [{\citenamefont {Vasiliev}\ \emph {et~al.}(2018)\citenamefont
  {Vasiliev}, \citenamefont {Volkova}, \citenamefont {Zvereva},\ and\
  \citenamefont {Markina}}]{Vasiliev2018}%
  \BibitemOpen
  \bibfield  {author} {\bibinfo {author} {\bibfnamefont {A.}~\bibnamefont
  {Vasiliev}}, \bibinfo {author} {\bibfnamefont {O.}~\bibnamefont {Volkova}},
  \bibinfo {author} {\bibfnamefont {E.}~\bibnamefont {Zvereva}}, \ and\
  \bibinfo {author} {\bibfnamefont {M.}~\bibnamefont {Markina}},\ }\href
  {\doibase 10.1038/s41535-018-0090-7} {\bibfield  {journal} {\bibinfo
  {journal} {npj Quantum Materials}\ }\textbf {\bibinfo {volume} {3}},\
  \bibinfo {pages} {18} (\bibinfo {year} {2018})}\BibitemShut {NoStop}%
\bibitem [{\citenamefont {Gardiner}\ and\ \citenamefont
  {Zoller}(2017)}]{Gardiner2017}%
  \BibitemOpen
  \bibfield  {author} {\bibinfo {author} {\bibfnamefont {C.}~\bibnamefont
  {Gardiner}}\ and\ \bibinfo {author} {\bibfnamefont {P.}~\bibnamefont
  {Zoller}},\ }\href {\doibase 10.1142/q0122} {\emph {\bibinfo {title} {The
  Quantum World of Ultra-Cold Atoms and Light Book {III}: Ultra-Cold Atoms}}}\
  (\bibinfo  {publisher} {{World} {Scientific} ({Europe})},\ \bibinfo {year}
  {2017})\BibitemShut {NoStop}%
\bibitem [{\citenamefont {Meher}\ and\ \citenamefont
  {Sivakumar}(2022)}]{Meher2022}%
  \BibitemOpen
  \bibfield  {author} {\bibinfo {author} {\bibfnamefont {N.}~\bibnamefont
  {Meher}}\ and\ \bibinfo {author} {\bibfnamefont {S.}~\bibnamefont
  {Sivakumar}},\ }\href@noop {} {\bibfield  {journal} {\bibinfo  {journal}
  {arXiv:2204.01322}\ } (\bibinfo {year} {2022})}\BibitemShut {NoStop}%
\bibitem [{\citenamefont {Makhlin}\ \emph {et~al.}(2001)\citenamefont
  {Makhlin}, \citenamefont {Sch\"on},\ and\ \citenamefont
  {Shnirman}}]{Makhlin2001}%
  \BibitemOpen
  \bibfield  {author} {\bibinfo {author} {\bibfnamefont {Y.}~\bibnamefont
  {Makhlin}}, \bibinfo {author} {\bibfnamefont {G.}~\bibnamefont {Sch\"on}}, \
  and\ \bibinfo {author} {\bibfnamefont {A.}~\bibnamefont {Shnirman}},\ }\href
  {\doibase 10.1103/RevModPhys.73.357} {\bibfield  {journal} {\bibinfo
  {journal} {Review Modern Physics}\ }\textbf {\bibinfo {volume} {73}},\
  \bibinfo {pages} {357} (\bibinfo {year} {2001})}\BibitemShut {NoStop}%
\bibitem [{\citenamefont {Luo}(2008)}]{Luo2008}%
  \BibitemOpen
  \bibfield  {author} {\bibinfo {author} {\bibfnamefont {S.~L.}\ \bibnamefont
  {Luo}},\ }\href {\doibase 10.1103/PhysRevA.77.042303} {\bibfield  {journal}
  {\bibinfo  {journal} {Physics Review A}\ }\textbf {\bibinfo {volume} {77}},\
  \bibinfo {pages} {042303} (\bibinfo {year} {2008})}\BibitemShut {NoStop}%
\bibitem [{\citenamefont {Kjaergaard}\ \emph {et~al.}(2020)\citenamefont
  {Kjaergaard}, \citenamefont {Schwartz}, \citenamefont {Braum\"{u}ller},
  \citenamefont {Krantz}, \citenamefont {Wang}, \citenamefont {Gustavsson},\
  and\ \citenamefont {Oliver}}]{Kjaergaard2020}%
  \BibitemOpen
  \bibfield  {author} {\bibinfo {author} {\bibfnamefont {M.}~\bibnamefont
  {Kjaergaard}}, \bibinfo {author} {\bibfnamefont {M.~E.}\ \bibnamefont
  {Schwartz}}, \bibinfo {author} {\bibfnamefont {J.}~\bibnamefont
  {Braum\"{u}ller}}, \bibinfo {author} {\bibfnamefont {P.}~\bibnamefont
  {Krantz}}, \bibinfo {author} {\bibfnamefont {J.~I.-J.}\ \bibnamefont {Wang}},
  \bibinfo {author} {\bibfnamefont {S.}~\bibnamefont {Gustavsson}}, \ and\
  \bibinfo {author} {\bibfnamefont {W.~D.}\ \bibnamefont {Oliver}},\ }\href
  {\doibase 10.1146/annurev-conmatphys-031119-050605} {\bibfield  {journal}
  {\bibinfo  {journal} {Annual Review of Condensed Matter Physics}\ }\textbf
  {\bibinfo {volume} {11}},\ \bibinfo {pages} {369} (\bibinfo {year}
  {2020})}\BibitemShut {NoStop}%
\bibitem [{\citenamefont {Penc}\ and\ \citenamefont
  {L{\"a}uchli}(2010)}]{Penc2010}%
  \BibitemOpen
  \bibfield  {author} {\bibinfo {author} {\bibfnamefont {K.}~\bibnamefont
  {Penc}}\ and\ \bibinfo {author} {\bibfnamefont {A.~M.}\ \bibnamefont
  {L{\"a}uchli}},\ }in\ \href {\doibase 10.1007/978-3-642-10589-0_13} {\emph
  {\bibinfo {booktitle} {Introduction to Frustrated Magnetism}}}\ (\bibinfo
  {publisher} {Springer Berlin Heidelberg},\ \bibinfo {year} {2010})\ pp.\
  \bibinfo {pages} {331--362}\BibitemShut {NoStop}%
\bibitem [{\citenamefont {Gutzwiller}(1963)}]{Gutzwiller1963}%
  \BibitemOpen
  \bibfield  {author} {\bibinfo {author} {\bibfnamefont {M.~C.}\ \bibnamefont
  {Gutzwiller}},\ }\href {\doibase 10.1103/physrevlett.10.159} {\bibfield
  {journal} {\bibinfo  {journal} {Physical Review Letters}\ }\textbf {\bibinfo
  {volume} {10}},\ \bibinfo {pages} {159} (\bibinfo {year} {1963})}\BibitemShut
  {NoStop}%
\bibitem [{\citenamefont {Gutzwiller}(1965)}]{Gutzwiller1965}%
  \BibitemOpen
  \bibfield  {author} {\bibinfo {author} {\bibfnamefont {M.~C.}\ \bibnamefont
  {Gutzwiller}},\ }\href {\doibase 10.1103/physrev.137.a1726} {\bibfield
  {journal} {\bibinfo  {journal} {Physical Review}\ }\textbf {\bibinfo {volume}
  {137}},\ \bibinfo {pages} {A1726} (\bibinfo {year} {1965})}\BibitemShut
  {NoStop}%
\bibitem [{\citenamefont {Cybenko}(1989)}]{Cybenko1989}%
  \BibitemOpen
  \bibfield  {author} {\bibinfo {author} {\bibfnamefont {G.}~\bibnamefont
  {Cybenko}},\ }\href {\doibase 10.1007/BF02551274} {\bibfield  {journal}
  {\bibinfo  {journal} {Mathematics of Control, Signals and Systems}\ }\textbf
  {\bibinfo {volume} {2}},\ \bibinfo {pages} {303} (\bibinfo {year}
  {1989})}\BibitemShut {NoStop}%
\bibitem [{\citenamefont {Hornik}\ \emph {et~al.}(1989)\citenamefont {Hornik},
  \citenamefont {Stinchcombe},\ and\ \citenamefont {White}}]{Hornik1989}%
  \BibitemOpen
  \bibfield  {author} {\bibinfo {author} {\bibfnamefont {K.}~\bibnamefont
  {Hornik}}, \bibinfo {author} {\bibfnamefont {M.}~\bibnamefont {Stinchcombe}},
  \ and\ \bibinfo {author} {\bibfnamefont {H.}~\bibnamefont {White}},\ }\href
  {\doibase 10.1016/0893-6080(89)90020-8} {\bibfield  {journal} {\bibinfo
  {journal} {Neural Networks}\ }\textbf {\bibinfo {volume} {2}},\ \bibinfo
  {pages} {359} (\bibinfo {year} {1989})}\BibitemShut {NoStop}%
\bibitem [{\citenamefont {Goodfellow}\ \emph {et~al.}(2016)\citenamefont
  {Goodfellow}, \citenamefont {Bengio},\ and\ \citenamefont
  {Courville}}]{Goodfellow2016}%
  \BibitemOpen
  \bibfield  {author} {\bibinfo {author} {\bibfnamefont {I.}~\bibnamefont
  {Goodfellow}}, \bibinfo {author} {\bibfnamefont {Y.}~\bibnamefont {Bengio}},
  \ and\ \bibinfo {author} {\bibfnamefont {A.}~\bibnamefont {Courville}},\
  }\href@noop {} {\emph {\bibinfo {title} {Deep Learning}}}\ (\bibinfo
  {publisher} {MIT Press},\ \bibinfo {year} {2016})\ \bibinfo {note}
  {\url{http://www.deeplearningbook.org}}\BibitemShut {NoStop}%
\bibitem [{\citenamefont {Iten}\ \emph {et~al.}(2020)\citenamefont {Iten},
  \citenamefont {Metger}, \citenamefont {Wilming}, \citenamefont {del Rio},\
  and\ \citenamefont {Renner}}]{Iten2020}%
  \BibitemOpen
  \bibfield  {author} {\bibinfo {author} {\bibfnamefont {R.}~\bibnamefont
  {Iten}}, \bibinfo {author} {\bibfnamefont {T.}~\bibnamefont {Metger}},
  \bibinfo {author} {\bibfnamefont {H.}~\bibnamefont {Wilming}}, \bibinfo
  {author} {\bibfnamefont {L.}~\bibnamefont {del Rio}}, \ and\ \bibinfo
  {author} {\bibfnamefont {R.}~\bibnamefont {Renner}},\ }\href {\doibase
  10.1103/physrevlett.124.010508} {\bibfield  {journal} {\bibinfo  {journal}
  {Physical Review Letters}\ }\textbf {\bibinfo {volume} {124}},\ \bibinfo
  {pages} {010508} (\bibinfo {year} {2020})}\BibitemShut {NoStop}%
\bibitem [{\citenamefont {Wang}\ \emph {et~al.}(2021)\citenamefont {Wang},
  \citenamefont {Wang}, \citenamefont {Zhang}, \citenamefont {Sun},\ and\
  \citenamefont {Xia}}]{Wang2021}%
  \BibitemOpen
  \bibfield  {author} {\bibinfo {author} {\bibfnamefont {W.}~\bibnamefont
  {Wang}}, \bibinfo {author} {\bibfnamefont {Z.}~\bibnamefont {Wang}}, \bibinfo
  {author} {\bibfnamefont {Y.}~\bibnamefont {Zhang}}, \bibinfo {author}
  {\bibfnamefont {B.}~\bibnamefont {Sun}}, \ and\ \bibinfo {author}
  {\bibfnamefont {K.}~\bibnamefont {Xia}},\ }\href {\doibase
  10.1103/physrevapplied.16.014005} {\bibfield  {journal} {\bibinfo  {journal}
  {Physical Review Applied}\ }\textbf {\bibinfo {volume} {16}},\ \bibinfo
  {pages} {014005} (\bibinfo {year} {2021})}\BibitemShut {NoStop}%
\bibitem [{\citenamefont {Rumelhart}\ \emph {et~al.}(1986)\citenamefont
  {Rumelhart}, \citenamefont {Hinton},\ and\ \citenamefont
  {Williams}}]{Rumelhart1986LIR}%
  \BibitemOpen
  \bibfield  {author} {\bibinfo {author} {\bibfnamefont {D.~E.}\ \bibnamefont
  {Rumelhart}}, \bibinfo {author} {\bibfnamefont {G.~E.}\ \bibnamefont
  {Hinton}}, \ and\ \bibinfo {author} {\bibfnamefont {R.~J.}\ \bibnamefont
  {Williams}},\ }in\ \href@noop {} {\emph {\bibinfo {booktitle} {Parallel
  Distributed Processing}}},\ \bibinfo {editor} {edited by\ \bibinfo {editor}
  {\bibfnamefont {D.~E.}\ \bibnamefont {Rumelhart}}\ and\ \bibinfo {editor}
  {\bibfnamefont {R.~J.}\ \bibnamefont {McClelland}}}\ (\bibinfo  {publisher}
  {MIT Press},\ \bibinfo {address} {Cambridge, Mass.},\ \bibinfo {year}
  {1986})\ Chap.~\bibinfo {chapter} {8}\BibitemShut {NoStop}%
\bibitem [{\citenamefont {P}\ and\ \citenamefont {Ba}(2014)}]{Adam2015}%
  \BibitemOpen
  \bibfield  {author} {\bibinfo {author} {\bibfnamefont {D.}~\bibnamefont {P},
  \bibfnamefont {Kingma}}\ and\ \bibinfo {author} {\bibfnamefont
  {J.}~\bibnamefont {Ba}},\ }\href@noop {} {\bibfield  {journal} {\bibinfo
  {journal} {arxiv:1412.6980}\ } (\bibinfo {year} {2014})}\BibitemShut
  {NoStop}%
\bibitem [{\citenamefont {Wilson}\ and\ \citenamefont
  {Martinez}(2003)}]{Wilson2003}%
  \BibitemOpen
  \bibfield  {author} {\bibinfo {author} {\bibfnamefont {D.}~\bibnamefont
  {Wilson}}\ and\ \bibinfo {author} {\bibfnamefont {T.~R.}\ \bibnamefont
  {Martinez}},\ }\href {\doibase 10.1016/s0893-6080(03)00138-2} {\bibfield
  {journal} {\bibinfo  {journal} {Neural Networks}\ }\textbf {\bibinfo {volume}
  {16}},\ \bibinfo {pages} {1429} (\bibinfo {year} {2003})}\BibitemShut
  {NoStop}%
\bibitem [{\citenamefont {Abadi}\ \emph {et~al.}(2016)\citenamefont {Abadi},
  \citenamefont {Barham}, \citenamefont {Chen}, \citenamefont {Chen},
  \citenamefont {Davis}, \citenamefont {Dean}, \citenamefont {Devin},
  \citenamefont {Ghemawat}, \citenamefont {Irving}, \citenamefont {Isard},
  \citenamefont {Kudlur}, \citenamefont {Levenberg}, \citenamefont {Monga},
  \citenamefont {Moore}, \citenamefont {Murray}, \citenamefont {Steiner},
  \citenamefont {Tucker}, \citenamefont {Vasudevan}, \citenamefont {Warden},
  \citenamefont {Wicke}, \citenamefont {Yu},\ and\ \citenamefont
  {Zheng}}]{Abadi2016}%
  \BibitemOpen
  \bibfield  {author} {\bibinfo {author} {\bibfnamefont {M.}~\bibnamefont
  {Abadi}}, \bibinfo {author} {\bibfnamefont {P.}~\bibnamefont {Barham}},
  \bibinfo {author} {\bibfnamefont {J.}~\bibnamefont {Chen}}, \bibinfo {author}
  {\bibfnamefont {Z.}~\bibnamefont {Chen}}, \bibinfo {author} {\bibfnamefont
  {A.}~\bibnamefont {Davis}}, \bibinfo {author} {\bibfnamefont
  {J.}~\bibnamefont {Dean}}, \bibinfo {author} {\bibfnamefont {M.}~\bibnamefont
  {Devin}}, \bibinfo {author} {\bibfnamefont {S.}~\bibnamefont {Ghemawat}},
  \bibinfo {author} {\bibfnamefont {G.}~\bibnamefont {Irving}}, \bibinfo
  {author} {\bibfnamefont {M.}~\bibnamefont {Isard}}, \bibinfo {author}
  {\bibfnamefont {M.}~\bibnamefont {Kudlur}}, \bibinfo {author} {\bibfnamefont
  {J.}~\bibnamefont {Levenberg}}, \bibinfo {author} {\bibfnamefont
  {R.}~\bibnamefont {Monga}}, \bibinfo {author} {\bibfnamefont
  {S.}~\bibnamefont {Moore}}, \bibinfo {author} {\bibfnamefont {D.~G.}\
  \bibnamefont {Murray}}, \bibinfo {author} {\bibfnamefont {B.}~\bibnamefont
  {Steiner}}, \bibinfo {author} {\bibfnamefont {P.}~\bibnamefont {Tucker}},
  \bibinfo {author} {\bibfnamefont {V.}~\bibnamefont {Vasudevan}}, \bibinfo
  {author} {\bibfnamefont {P.}~\bibnamefont {Warden}}, \bibinfo {author}
  {\bibfnamefont {M.}~\bibnamefont {Wicke}}, \bibinfo {author} {\bibfnamefont
  {Y.}~\bibnamefont {Yu}}, \ and\ \bibinfo {author} {\bibfnamefont
  {X.}~\bibnamefont {Zheng}},\ }in\ \href
  {https://www.usenix.org/conference/osdi16/technical-sessions/presentation/abadi}
  {\emph {\bibinfo {booktitle} {12th {USENIX} Symposium on Operating Systems
  Design and Implementation ({OSDI} 16)}}}\ (\bibinfo  {publisher} {{USENIX}
  Association},\ \bibinfo {address} {Savannah, GA},\ \bibinfo {year} {2016})\
  pp.\ \bibinfo {pages} {265--283}\BibitemShut {NoStop}%
\bibitem [{\citenamefont {Tomonaga}(1950)}]{Tomonaga1950}%
  \BibitemOpen
  \bibfield  {author} {\bibinfo {author} {\bibfnamefont {S.~I.}\ \bibnamefont
  {Tomonaga}},\ }\href {\doibase 10.1143/ptp/5.4.544} {\bibfield  {journal}
  {\bibinfo  {journal} {Progress of Theoretical Physics}\ }\textbf {\bibinfo
  {volume} {5}},\ \bibinfo {pages} {544} (\bibinfo {year} {1950})}\BibitemShut
  {NoStop}%
\bibitem [{\citenamefont {Luttinger}(1963)}]{Luttinger1963}%
  \BibitemOpen
  \bibfield  {author} {\bibinfo {author} {\bibfnamefont {J.~M.}\ \bibnamefont
  {Luttinger}},\ }\href {\doibase 10.1063/1.1704046} {\bibfield  {journal}
  {\bibinfo  {journal} {Journal of Mathematical Physics}\ }\textbf {\bibinfo
  {volume} {4}},\ \bibinfo {pages} {1154} (\bibinfo {year} {1963})}\BibitemShut
  {NoStop}%
\bibitem [{\citenamefont {Haldane}(1981)}]{Haldane1981}%
  \BibitemOpen
  \bibfield  {author} {\bibinfo {author} {\bibfnamefont {F.~D.~M.}\
  \bibnamefont {Haldane}},\ }\href {\doibase 10.1088/0022-3719/14/19/010}
  {\bibfield  {journal} {\bibinfo  {journal} {Journal of Physics C: Solid State
  Physics}\ }\textbf {\bibinfo {volume} {14}},\ \bibinfo {pages} {2585}
  (\bibinfo {year} {1981})}\BibitemShut {NoStop}%
\bibitem [{\citenamefont {White}\ and\ \citenamefont
  {Affleck}(1996)}]{White1996}%
  \BibitemOpen
  \bibfield  {author} {\bibinfo {author} {\bibfnamefont {S.~R.}\ \bibnamefont
  {White}}\ and\ \bibinfo {author} {\bibfnamefont {I.}~\bibnamefont
  {Affleck}},\ }\href {\doibase 10.1103/physrevb.54.9862} {\bibfield  {journal}
  {\bibinfo  {journal} {Physical Review B}\ }\textbf {\bibinfo {volume} {54}},\
  \bibinfo {pages} {9862} (\bibinfo {year} {1996})}\BibitemShut {NoStop}%
\bibitem [{\citenamefont {Soos}\ \emph {et~al.}(2016)\citenamefont {Soos},
  \citenamefont {Parvej},\ and\ \citenamefont {Kumar}}]{Soos2016}%
  \BibitemOpen
  \bibfield  {author} {\bibinfo {author} {\bibfnamefont {Z.~G.}\ \bibnamefont
  {Soos}}, \bibinfo {author} {\bibfnamefont {A.}~\bibnamefont {Parvej}}, \ and\
  \bibinfo {author} {\bibfnamefont {M.}~\bibnamefont {Kumar}},\ }\href
  {\doibase 10.1088/0953-8984/28/17/175603} {\bibfield  {journal} {\bibinfo
  {journal} {Journal of Physics: Condensed Matter}\ }\textbf {\bibinfo {volume}
  {28}},\ \bibinfo {pages} {175603} (\bibinfo {year} {2016})}\BibitemShut
  {NoStop}%
\bibitem [{\citenamefont {Anderson}(1967)}]{Anderson1967}%
  \BibitemOpen
  \bibfield  {author} {\bibinfo {author} {\bibfnamefont {P.~W.}\ \bibnamefont
  {Anderson}},\ }\href {\doibase 10.1103/physrevlett.18.1049} {\bibfield
  {journal} {\bibinfo  {journal} {Physical Review Letters}\ }\textbf {\bibinfo
  {volume} {18}},\ \bibinfo {pages} {1049} (\bibinfo {year}
  {1967})}\BibitemShut {NoStop}%
\bibitem [{\citenamefont {Gu}(2010)}]{Gu2010}%
  \BibitemOpen
  \bibfield  {author} {\bibinfo {author} {\bibfnamefont {S.~J.}\ \bibnamefont
  {Gu}},\ }\href {\doibase 10.1142/s0217979210056335} {\bibfield  {journal}
  {\bibinfo  {journal} {International Journal of Modern Physics B}\ }\textbf
  {\bibinfo {volume} {24}},\ \bibinfo {pages} {4371} (\bibinfo {year}
  {2010})}\BibitemShut {NoStop}%
\bibitem [{\citenamefont {Wang}\ \emph {et~al.}(2009)\citenamefont {Wang},
  \citenamefont {Gu},\ and\ \citenamefont {Chen}}]{Wang2009}%
  \BibitemOpen
  \bibfield  {author} {\bibinfo {author} {\bibfnamefont {L.}~\bibnamefont
  {Wang}}, \bibinfo {author} {\bibfnamefont {S.-J.}\ \bibnamefont {Gu}}, \ and\
  \bibinfo {author} {\bibfnamefont {S.}~\bibnamefont {Chen}},\ }\href@noop {}
  {\bibfield  {journal} {\bibinfo  {journal} {arXiv:0903.4242}\ } (\bibinfo
  {year} {2009})}\BibitemShut {NoStop}%
\bibitem [{\citenamefont {Cincio}\ \emph {et~al.}(2019)\citenamefont {Cincio},
  \citenamefont {Rams}, \citenamefont {Dziarmaga},\ and\ \citenamefont
  {Zurek}}]{Cincio2019}%
  \BibitemOpen
  \bibfield  {author} {\bibinfo {author} {\bibfnamefont {L.}~\bibnamefont
  {Cincio}}, \bibinfo {author} {\bibfnamefont {M.~M.}\ \bibnamefont {Rams}},
  \bibinfo {author} {\bibfnamefont {J.}~\bibnamefont {Dziarmaga}}, \ and\
  \bibinfo {author} {\bibfnamefont {W.~H.}\ \bibnamefont {Zurek}},\ }\href
  {\doibase 10.1103/physrevb.100.081108} {\bibfield  {journal} {\bibinfo
  {journal} {Physical Review B}\ }\textbf {\bibinfo {volume} {100}},\ \bibinfo
  {pages} {081108(R)} (\bibinfo {year} {2019})}\BibitemShut {NoStop}%
\bibitem [{\citenamefont {Bach}\ \emph {et~al.}(2021)\citenamefont {Bach},
  \citenamefont {Cicalese}, \citenamefont {Kreutz},\ and\ \citenamefont
  {Orlando}}]{Bach2021}%
  \BibitemOpen
  \bibfield  {author} {\bibinfo {author} {\bibfnamefont {A.}~\bibnamefont
  {Bach}}, \bibinfo {author} {\bibfnamefont {M.}~\bibnamefont {Cicalese}},
  \bibinfo {author} {\bibfnamefont {L.}~\bibnamefont {Kreutz}}, \ and\ \bibinfo
  {author} {\bibfnamefont {G.}~\bibnamefont {Orlando}},\ }\href {\doibase
  10.1007/s00526-021-02016-3} {\bibfield  {journal} {\bibinfo  {journal}
  {Calculus of Variations and Partial Differential Equations}\ }\textbf
  {\bibinfo {volume} {60}},\ \bibinfo {pages} {149} (\bibinfo {year}
  {2021})}\BibitemShut {NoStop}%
\bibitem [{\citenamefont {Wang}\ \emph {et~al.}(2018)\citenamefont {Wang},
  \citenamefont {Wang},\ and\ \citenamefont {Sedrakyan}}]{Wang2018}%
  \BibitemOpen
  \bibfield  {author} {\bibinfo {author} {\bibfnamefont {R.}~\bibnamefont
  {Wang}}, \bibinfo {author} {\bibfnamefont {B.}~\bibnamefont {Wang}}, \ and\
  \bibinfo {author} {\bibfnamefont {T.~A.}\ \bibnamefont {Sedrakyan}},\ }\href
  {\doibase 10.1103/physrevb.98.064402} {\bibfield  {journal} {\bibinfo
  {journal} {Physical Review B}\ }\textbf {\bibinfo {volume} {98}},\ \bibinfo
  {pages} {064402} (\bibinfo {year} {2018})}\BibitemShut {NoStop}%
\bibitem [{\citenamefont {Sedrakyan}\ \emph {et~al.}(2020)\citenamefont
  {Sedrakyan}, \citenamefont {Moessner},\ and\ \citenamefont
  {Kamenev}}]{Sedrakyan2020}%
  \BibitemOpen
  \bibfield  {author} {\bibinfo {author} {\bibfnamefont {T.}~\bibnamefont
  {Sedrakyan}}, \bibinfo {author} {\bibfnamefont {R.}~\bibnamefont {Moessner}},
  \ and\ \bibinfo {author} {\bibfnamefont {A.}~\bibnamefont {Kamenev}},\ }\href
  {\doibase 10.1103/physrevb.102.024430} {\bibfield  {journal} {\bibinfo
  {journal} {Physical Review B}\ }\textbf {\bibinfo {volume} {102}},\ \bibinfo
  {pages} {024430} (\bibinfo {year} {2020})}\BibitemShut {NoStop}%
\bibitem [{\citenamefont {Henderson}\ \emph {et~al.}(1992)\citenamefont
  {Henderson}, \citenamefont {Oitmaa},\ and\ \citenamefont
  {Ashley}}]{Henderson1992}%
  \BibitemOpen
  \bibfield  {author} {\bibinfo {author} {\bibfnamefont {J.~A.}\ \bibnamefont
  {Henderson}}, \bibinfo {author} {\bibfnamefont {J.}~\bibnamefont {Oitmaa}}, \
  and\ \bibinfo {author} {\bibfnamefont {M.~C.~B.}\ \bibnamefont {Ashley}},\
  }\href {\doibase 10.1103/PhysRevB.46.6328} {\bibfield  {journal} {\bibinfo
  {journal} {Phys. Rev. B}\ }\textbf {\bibinfo {volume} {46}},\ \bibinfo
  {pages} {6328} (\bibinfo {year} {1992})}\BibitemShut {NoStop}%
\bibitem [{\citenamefont {Essler}\ \emph {et~al.}(2005)\citenamefont {Essler},
  \citenamefont {Frahm}, \citenamefont {G{\"o}hmann}, \citenamefont
  {Kl{\"u}mper},\ and\ \citenamefont {Korepin}}]{Essler2005}%
  \BibitemOpen
  \bibfield  {author} {\bibinfo {author} {\bibfnamefont {F.~H.}\ \bibnamefont
  {Essler}}, \bibinfo {author} {\bibfnamefont {H.}~\bibnamefont {Frahm}},
  \bibinfo {author} {\bibfnamefont {F.}~\bibnamefont {G{\"o}hmann}}, \bibinfo
  {author} {\bibfnamefont {A.}~\bibnamefont {Kl{\"u}mper}}, \ and\ \bibinfo
  {author} {\bibfnamefont {V.~E.}\ \bibnamefont {Korepin}},\ }\href@noop {}
  {\emph {\bibinfo {title} {The one-dimensional Hubbard model}}}\ (\bibinfo
  {publisher} {Cambridge University Press},\ \bibinfo {year}
  {2005})\BibitemShut {NoStop}%
\bibitem [{\citenamefont {Moriya}(2012)}]{Moriya2012}%
  \BibitemOpen
  \bibfield  {author} {\bibinfo {author} {\bibfnamefont {T.}~\bibnamefont
  {Moriya}},\ }\href@noop {} {\emph {\bibinfo {title} {Electron Correlation and
  Magnetism in Narrow-Band Systems: Proceedings of the Third Taniguchi
  International Symposium, Mount Fuji, Japan, November 1--5, 1980}}},\
  Vol.~\bibinfo {volume} {29}\ (\bibinfo  {publisher} {Springer Science \&
  Business Media},\ \bibinfo {year} {2012})\BibitemShut {NoStop}%
\bibitem [{\citenamefont {Derzhko}(2001)}]{Derzhko2001}%
  \BibitemOpen
  \bibfield  {author} {\bibinfo {author} {\bibfnamefont {O.}~\bibnamefont
  {Derzhko}},\ }\href@noop {} {\bibfield  {journal} {\bibinfo  {journal}
  {arXiv:cond-mat/0101188}\ } (\bibinfo {year} {2001})}\BibitemShut {NoStop}%
\bibitem [{\citenamefont {Lieb}\ and\ \citenamefont {Wu}(1968)}]{Lieb1968}%
  \BibitemOpen
  \bibfield  {author} {\bibinfo {author} {\bibfnamefont {E.~H.}\ \bibnamefont
  {Lieb}}\ and\ \bibinfo {author} {\bibfnamefont {F.~Y.}\ \bibnamefont {Wu}},\
  }\href {\doibase 10.1103/physrevlett.20.1445} {\bibfield  {journal} {\bibinfo
   {journal} {Physical Review Letters}\ }\textbf {\bibinfo {volume} {20}},\
  \bibinfo {pages} {1445} (\bibinfo {year} {1968})}\BibitemShut {NoStop}%
\end{thebibliography}%

\end{document}